\documentclass[11pt]{article}

\include{setup}

\begin{document}

\thispagestyle{empty}
\def\thefootnote{\fnsymbol{footnote}}
\setcounter{footnote}{1}
\null
\draftdate\hfill MPP-2009-72 \\
\strut\hfill PSI-PR-09-08 \\
\vskip 0cm
\vfill
\begin{center}
  {\Large \boldmath{\bf Electroweak corrections to $\PW+\mathrm{jet}$
      hadroproduction \\[.5em] including leptonic W-boson decays}
\par} \vskip 2.5em
{\large
{\sc Ansgar Denner$^{1}$, Stefan Dittmaier$^{2,3}$, 
     Tobias Kasprzik$^{3}$, Alexander M\"uck$^{1}$
}\\[2ex]
{\normalsize \it 
$^1$Paul Scherrer Institut, W\"urenlingen und Villigen,\\ 
Ch-5232 Villigen PSI, Switzerland}
\\[1ex]
{\normalsize \it 
$^2$Albert-Ludwigs-Universit\"at Freiburg, Physikalisches Institut, \\
D-79104 Freiburg, Germany
}\\[1ex]
{\normalsize \it 
$^3$Max-Planck-Institut f\"ur Physik (Werner-Heisenberg-Institut), \\
D-80805 M\"unchen, Germany
}\\[2ex]
}
\par \vskip 1em
\end{center}\par
\vskip .0cm \vfill {\bf Abstract:} \par 
We present the first calculation of the next-to-leading-order
electroweak corrections to W-boson + jet hadroproduction including
leptonic W-boson decays.  The W-boson resonance is treated
consistently using the complex-mass scheme, and all off-shell effects
are taken into account. The corresponding next-to-leading-order QCD
corrections have also been recalculated. All the results are
implemented in a flexible Monte Carlo code. Numerical results for
cross sections and distributions of this Standard Model benchmark
process are presented for the Tevatron and the LHC.

\par
\vskip 1cm
\noindent
August 2009
\par
\null
\setcounter{page}{0}
\clearpage
\def\thefootnote{\arabic{footnote}}
\setcounter{footnote}{0}

\section{Introduction}

The production of electroweak (EW) \PW\ and \PZ\ bosons with subsequent
leptonic decays is one of the most prominent Standard Model (SM)
processes at present and future hadron colliders like the Tevatron and
the LHC. The signatures are clean owing to the final-state leptons, and
the cross sections are large. 
The (expected) experimental accuracy is so excellent that the charged-current 
Drell--Yan process allows to improve the precision
measurement of the \PW-boson mass.
Moreover, it can deliver important constraints in the fit of the
parton distribution functions (PDFs) and may serve as
a luminosity monitor at the LHC.
The off-shell tails of appropriate
distributions give access to a \PW-width measurement and, at high
energies, offer the possibility to search for new charged
$\PW{}^\prime$ gauge bosons. (See e.g.\ 
\citere{Gerber:2007xk,Buescher:2006jm} and references therein.)

At hadron colliders, the EW gauge bosons are (almost) always
produced together with additional QCD radiation. The production
cross section of \PW\ bosons in association with a hard, visible jet,
\beq
\Pp\Pp/\Pp\bar\Pp \to \PW  + \mathrm{jet} \to \Pl\nu_\Pl + \mathrm{jet} +\X,
\eeq
is still large. Moreover, the intermediate \PW\ boson recoils against
the jet leading to a new kinematical situation with strongly boosted
\PW\ bosons.  For large transverse momentum ($p_{\rT}$) of the jet the
corresponding events contain charged leptons and/or neutrinos with
large $p_{\rT}$. In fact, in the SM, $\PW+\mathrm{jet(s)}$ production
is the largest source for events with large missing transverse
momentum where also a charged lepton is present for triggering. Hence,
$\PW+\mathrm{jet(s)}$ production is not only a SM candle process. It
is also an important background for a large class of new physics
searches based on missing transverse momentum. Moreover, the process
offers the possibility for precision tests concerning jet dynamics in
QCD.

To match the prospects and importance of this process class, an
excellent theoretical prediction is mandatory. The differential cross
section for \PW-boson production is known at NNLO accuracy with
respect to QCD corrections~\cite{vanNeerven:1991gh} and even up to
N$^3$LO in the soft-plus-virtual approximation~\cite{Moch:2005ky}. The
next-to-leading-order (NLO) QCD corrections have been matched with
parton showers \cite{Frixione:2006gn} and combined with a summation of
soft-gluon radiation (see e.g.\ \citere{Balazs:1997xd}), which is
particularly important to reliably predict the
transverse-momentum distribution of the \PW\ bosons for small
$p_{\rT}$. A theoretical study of the QCD uncertainties in the
determination of the \PW~cross section at hadron colliders has been
presented in \citere{Frixione:2004us}. Concerning EW corrections, the
full NLO~\cite{Zykunov:2001mn, Dittmaier:2001ay, Baur:2004ig,
  CarloniCalame:2006zq} and leading higher-order effects, in
particular due to multi-photon final-state
radiation~\cite{CarloniCalame:2003ux,Placzek:2003zg,
  CarloniCalame:2006zq,Brensing:2007qm}, have been calculated. The
contributions of photon-induced processes have been discussed in
\citeres{DKLH,Arbuzov:2007kp,Brensing:2007qm}.  First steps towards
combining QCD and EW higher-order effects have been taken in
\citere{Cao:2004yy}. The NLO QCD and EW corrections have been also
calculated within the MSSM~\cite{Brensing:2007qm}.

The cross section for $\PW+1\,\mathrm{jet}$~\cite{Giele:1993dj,Campbell:2002tg} 
and $\PW+2\,\mathrm{jets}$~\cite{Campbell:2002tg} production is known at NLO
QCD. The calculation of the NLO QCD corrections in the leading-colour
approximation for the $\PW+3\,\mathrm{jets}$ cross section has recently been 
completed~\cite{Ellis:2009zw}. 

So far, the EW corrections in the SM have been assessed for $\PW+1\,\mathrm{jet}$
production in the approximation where the \PW\ boson is treated as a
stable external particle~\cite{Kuhn:2007qc,Kuhn:2007cv,Hollik:2007sq}
(see \citere{Gounaris:2007gx} for an MSSM analysis). For \PW\ bosons
at large transverse momentum, i.e.\ at large centre-of-mass energy,
this is a good approximation since the EW corrections are dominated by
large universal Sudakov logarithms \cite{Ciafaloni:1998xg}.
However, an on-shell calculation cannot assess any off-shell effects
due to the finite width of the \PW\ boson and is blind to the
details of the experimental event selection based on the charged-lepton 
momentum and the missing transverse momentum of the neutrino.

In this work, we present a calculation of the NLO EW
corrections for the physical final state in \PW-boson hadroproduction,
i.e.\ $\Pp\Pp/\Pp\bar\Pp \to \Pl\nu_\Pl + \mathrm{jet} +\X$. The \PW\ 
resonance is described in the complex-mass 
scheme~\cite{Denner:1999gp,Denner:2005fg}. All off-shell
effects due to the finite width of the \PW\ boson are included. Our
results have been implemented in a fully flexible Monte Carlo code
which is able to calculate binned distributions for all physically
relevant $\PW+1\,\mathrm{jet}$ observables.  In real emission events with
photons inside a jet, we distinguish $\PW+\mathrm{jet}$ and $\PW+\mathrm{photon}$
production by a cut on the photon energy fraction inside the jet
employing a measured quark-to-photon fragmentation function
\cite{Buskulic:1995au}.

Our calculation is completely generic in the sense that it can
predict observables which are dominated by \PW\ bosons close to their
mass shell as well as observables for which the exchanged \PW\ boson
is far off-shell.  The calculation of the EW corrections for \PW\ 
production in association with a hard jet is also a step towards a
better understanding of the interplay between QCD and EW corrections
for \PW\ production in general. This understanding---including a full
treatment of off-shell W~bosons---is mandatory to
match the envisaged experimental accuracy for the \PW-mass measurement
at the Tevatron and the LHC.

To reach the accuracy of $\mathcal{O}(\alpha_\mathrm{s} \alpha^3)$
throughout the calculation we have also included the photon-induced
partonic processes and the respective NLO QCD corrections. Also
non-trivial interference terms between EW and QCD diagrams
within the real corrections have been
included at this order. Moreover, we have recalculated the NLO QCD
corrections at $\mathcal{O}(\alpha^2_\mathrm{s} \alpha^2)$ in a fully
flexible way, supporting a phase-space dependent choice for the
factorization and renormalization scales.

This paper is organized as follows. In \refse{se:details}, we describe
our calculation in detail and discuss all the theoretical concepts and
tools which have been used. In \refse{se:numres}, we specify the
numerical input as well as the details of our event selection.
Numerical results are given for $\PW^+$ production both at the LHC and
at the Tevatron. We present inclusive cross sections for specified sets
of cuts as well as distributions for the relevant observables. We
conclude in \refse{se:concl}.

\section{Details of the calculation}
\label{se:details} 

\subsection{General setup}
\label{se:setup} 

The hadroproduction of a \PW\ boson in association with one hard jet
is governed at leading order (LO) by quark--antiquark fusion, where the
initial-state quarks radiate a gluon, and the corresponding crossed
channels with a gluon in the initial state. 
Specifically, for \PWp\ production the relevant partonic
processes are
\begin{eqnarray}
\label{eq:proc1}
& \Pu_i \, \, \Pdbar_j & \to \PWp \Pg      \to \Plp \Pnl \, \Pg \, ,\\
\label{eq:proc2}
& \Pu_i \, \, \Pg      & \to \PWp \Pd_j    \to \Plp \Pnl \, \Pd_j \, ,\\
\label{eq:proc3}
& \Pdbar_j \, \, \Pg   & \to \PWp \Pubar_i \to \Plp \Pnl \, \Pubar_i \, ,
\end{eqnarray}
where $\Pu_i$ and $\Pd_j$ denote an up-type quark of generation $i$
and a down-type quark of generation $j$, respectively.  We perform the
calculation for the physical final state, i.e.\ a charged lepton \Pl,
the corresponding neutrino \Pnl, and a parton which will be seen in
the detector as a jet. The corresponding tree-level Feynman diagrams
for process \refeq{eq:proc2} are shown in \reffi{fi:born_udWg}. The
intermediate \PW-boson resonance is described by a complex \PW-boson
mass $ \mu_{\PW}$ via the replacement
\bfi[t]
\begin{center}
\vspace*{1em}
\includegraphics{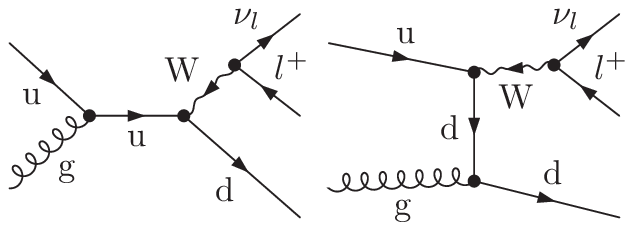}
\end{center}
\mycaption{\label{fi:born_udWg} Feynman diagrams for the LO process \refeq{eq:proc2}.}
\efi
\begin{equation}
\MW^2 \to \mu_{\PW}^2 = \MW^2 -\ri \MW \GW
\end{equation}
in the \PW~propagator as dictated by the complex-mass scheme (see
below). Hence, all our results correspond to a fixed-width description
of the Breit--Wigner resonance. The leptons are treated as massless
unless their small masses are used to regularize a collinear
divergence.

The dependence on quark mixing, as parametrized in the CKM matrix,
factorizes from the tree-level matrix elements. Apart from a global
CKM-dependent factor, the tree-level amplitudes do not depend on the
specific flavours. Hence, for hadronic observables, the summation over
the quark flavours $i=1,2$ and $j=1,2,3$ requires only the evaluation
of a single generic amplitude per process type shown in
\refeq{eq:proc1}--\refeq{eq:proc3} when
folding the squared tree-level amplitudes with the corresponding
PDFs. Only squares of the absolute
value of CKM elements enter the final results. We do not include
top quarks in the final state since their decays lead to significantly
different signatures. The five other quark flavours (including the
bottom quark) are treated as massless throughout the calculation, 
except if small masses are needed to regularize a collinear
divergence. Since we neglect the small CKM mixing of the third
generation with the first two generations, the PDFs of the bottom
quark are irrelevant at tree level but enter the result for the QCD
bremsstrahlung cross sections (see \refse{se:real}).

In this work, we describe $\PW+\mathrm{jet}$ production up to an accuracy of
$\mathcal{O}(\alpha^3 \alpha_{\mathrm{s}})$. Hence, we also include the 
$\mathcal{O}(\alpha^3)$ tree-level processes with a photon in the initial state,
\begin{eqnarray}
\label{eq:proc4}
& \Pu_i \, \, \ga      & \to \PWp \Pd_j    \to \Plp \Pnl \, \Pd_j \, ,\\
\label{eq:proc5}
& \Pdbar_j \, \, \ga   & \to \PWp \Pubar_i \to \Plp \Pnl \, \Pubar_i \, .
\end{eqnarray}
\bfi[t]
\begin{center}
\vspace*{1em}
\includegraphics{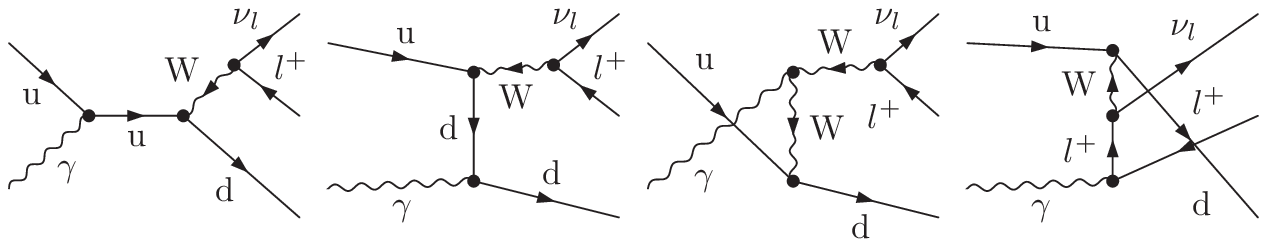}
\end{center}
\mycaption{\label{fi:born_ugaWd} 
Feynman diagrams for the photon-induced process \refeq{eq:proc4}.}
\efi
The tree-level Feynman diagrams for process \refeq{eq:proc4} are shown
in \reffi{fi:born_ugaWd}. The photon content of the proton has been
quantified in the MRSTQED2004 PDFs~\cite{Martin:2004dh}. 
Since the photon also couples to
the charged lepton and the intermediate \PW\ boson, the amplitude is
more involved than its QCD counterpart. In this work, we do not
consider the crossed processes corresponding to $\PW+\mathrm{photon}$
production. At tree level, $\PW+\mathrm{jet}$ and
$\PW+\mathrm{photon}$ final states can be distinguished trivially,
however, at NLO the definition of the $\PW+\mathrm{jet}$ final state
has to be done with care when additional photons are present.  This
issue and our treatment are discussed in detail in \refse{se:real}.

To define the electromagnetic coupling constant $\alpha$, 
we use the $\GF$ scheme, \ie we derive $\alpha$ from the Fermi constant according to 
\beq
 \alpha_{\GF} = \frac{\sqrt{2}\GF\MW^2}{\pi}\left(1-\frac{\MW^2}{\MZ^2}\right).
\eeq
In this scheme, the weak corrections to muon decay $\De r$ are
included in the charge renormalization constant (see \eg
\citere{Dittmaier:2001ay}).  As a consequence, the EW corrections are
independent of logarithms of the light-quark masses. Moreover, this
definition effectively resums the contributions associated with the
running of $\al$ from zero to the W-boson mass and absorbs leading
universal corrections $\propto\GF\Mt^2$ from the $\rho$~parameter into
the LO amplitude.

For corrections due to collinear final-state radiation it would be
more appropriate to use $\alpha(0)$ defined in the Thomson limit to
describe the corresponding coupling. On the other hand, using
$\alpha_{\GF}$ everywhere is best suited to describe the large
corrections due to Sudakov logarithms in the high-energy regime. Thus, the
optimal choice cannot be achieved in one particular input
scheme and necessarily requires more refinements. In particular,
among other things, higher-order effects from multi-photon emission
should also be included at this level of precision which is beyond
the scope of this work. The difference of the two schemes only
amounts to about 3\% of the EW corrections.

We employ the traditional Feynman-diagrammatic approach to calculate all relevant
amplitudes in the 't Hooft--Feynman gauge. For a numerical evaluation at the amplitude
level we use the Weyl--van-der-Waerden spinor formalism. 
To ensure the correctness of the presented results we have performed two 
independent calculations which are in mutual agreement. 

One calculation starts from diagrammatic expressions for the one-loop
corrections generated by {\sc FeynArts} 1.0 \cite{Kublbeck:1990xc}.
The algebraic evaluation of the loop amplitudes is performed with an 
in-house program written in {\sl Mathematica}, and the results are 
automatically transferred to {\sl Fortran}. 
The Born and bremsstrahlung amplitudes are
calculated and optimised by hand and directly included into a {\sl Fortran}
program for numerical evaluation. A specific parametrization of phase
space is used for an adaptive Monte Carlo integration employing the
{\sc Vegas} \cite{Lepage:1977sw} algorithm.

The second calculation is based on {\sc FeynArts} 3.2
\cite{Hahn:2000kx} and {\sc FormCalc} version 3.1 \cite{Hahn:1998yk}.
The translation of the amplitudes into the Weyl--van-der-Waerden
formalism as presented in \citere{Dittmaier:1998nn} is performed with
the program {\sc Pole}~\cite{Accomando:2005ra}.  {\sc Pole} also
provides an interface to the multi-channel phase-space integrator {\sc
  Lusifer} \cite{Dittmaier:2002ap} which has been extended to use {\sc
  Vegas}~\cite{Lepage:1977sw} in order to optimise each phase-space
mapping.

\subsection{Virtual corrections}
\label{se:virt} 

We calculate the virtual one-loop QCD and EW corrections
for the partonic processes \refeq{eq:proc1}--\refeq{eq:proc3} to order
$\mathcal{O}(\alpha^2 \alpha_{\mathrm{s}}^2)$ and 
$\mathcal{O}(\alpha^3 \alpha_{\mathrm{s}})$,
respectively. Since the partonic processes
\refeq{eq:proc4} and \refeq{eq:proc5} are already suppressed by
$\alpha/\alpha_{\mathrm{s}}$ at LO, we only need to include the NLO QCD
corrections for these channels to reach the required accuracy.  The
QCD corrections are straight-forward to implement and are induced by
self-energy, vertex, and box (4-point) diagrams only. The NLO EW
corrections are more involved and include pentagon (5-point) diagrams.
There are $\mathcal{O}(100)$ diagrams per partonic channel, including
6 pentagons and 20 boxes.  The generic structure of the contributing
diagrams is indicated in \reffi{fi:EW_VF_udWg}, and the pentagon
diagrams are explicitly given in \reffi{fi:EW_pent_udWg}.  The
different channels are related by crossing symmetry.

\bfi
\begin{center}
\includegraphics{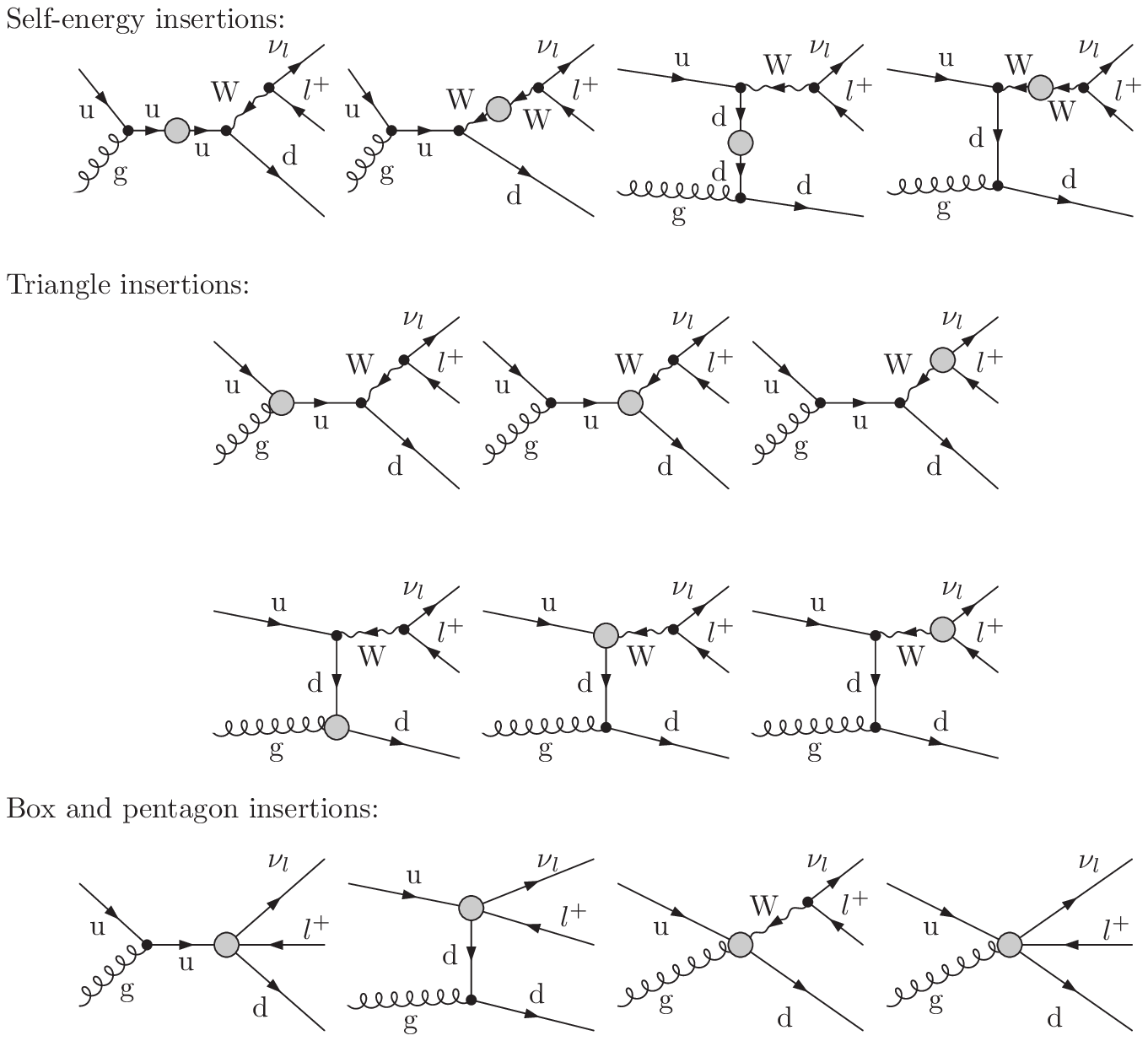}
\end{center}
\mycaption{\label{fi:EW_VF_udWg} Contributions of different
  one-particle irreducible vertex functions (indicated as blobs)
  to the LO process
  \refeq{eq:proc2}; there are contributions from self-energies,
  triangles, boxes, and pentagon graphs.}
\efi

\bfi
\begin{center}
\includegraphics{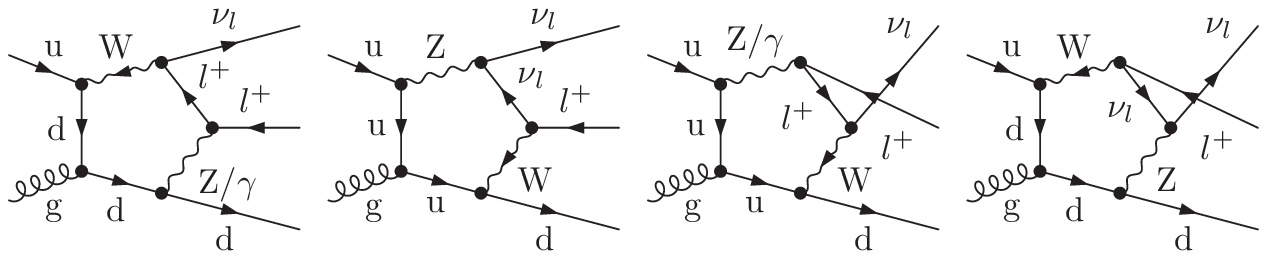}
\end{center}
\mycaption{\label{fi:EW_pent_udWg}
Virtual pentagon contributions to the process \refeq{eq:proc2}.}
\efi

The potentially resonant \PW\ bosons require a proper inclusion of the
finite gauge-boson width in the propagators. We use the complex-mass
scheme, which was introduced in \citere{Denner:1999gp} for LO
calculations and generalized to the one-loop level in
\citere{Denner:2005fg}. In this approach the W- and Z-boson masses are
consistently considered as complex quantities, defined as the
locations of the propagator poles in the complex plane.  This leads to
complex couplings and, in particular, a complex weak mixing angle.
The scheme fully respects all relations that follow from gauge
invariance. A brief description of the complex-mass scheme can also be
found in \citere{Denner:2006ic}.

The amplitudes can be expressed in terms of standard matrix elements
and coefficients, which contain the tensor integrals (following the
ideas in the appendix of \citere{Denner:2003iy}).  The tensor
integrals are recursively reduced to master integrals at the numerical
level. The standard scalar integrals are evaluated for complex masses
based on the methods and results of \citere{'tHooft:1979xw} using two
independent {\sl Fortran} implementations which are in mutual agreement.
Results for different regularization schemes are translated into each
other with the method of \citere{Dittmaier:2003bc}.  Tensor and scalar
5-point functions are directly expressed in terms of 4-point integrals
\cite{Melrose:1965kb,Denner:2002ii,Denner:2005nn}. Tensor 4-point and 3-point
integrals are reduced to scalar integrals with the Passarino--Veltman
algorithm \cite{Passarino:1979jh}. Although we already find sufficient
numerical stability with this procedure, we apply the dedicated
expansion methods of \citere{Denner:2005nn} in exceptional phase-space
regions where small Gram determinants appear.

UV divergences are regularized dimensionally. For the infrared (IR),
i.e.\ soft or collinear, divergences we either use pure dimensional
regularization with massless gluons, photons, and fermions (except for the
top quark), or
pure mass regularization with infinitesimal photon, gluon, and small
fermion masses, which are only kept in the
mass-singular logarithms.  When using dimensional regularization, the
rational terms of IR origin are treated as described in Appendix~A of
\citere{Bredenstein:2008zb}.

We use an on-shell renormalization prescription for the EW
part of the SM as detailed in \citere{Denner:2005fg} for the 
complex-mass scheme. Employing the \GF\ scheme for the definition of the 
fine-structure constant, we include $\De r$ in the charge renormalization
constant as mentioned above. The strong coupling constant is
renormalized in the \MSbar\ scheme with five active flavours. Hence,
bottom quarks are included everywhere in the calculation as a massless
quark flavour.

\subsection{Real corrections}
\label{se:real} 

The evaluation of the real corrections has to be done with 
particular care, both for theoretical
consistency as well as to match the experimental observables as closely as 
possible.
Let us first focus on the EW real corrections to the partonic processes
\refeq{eq:proc1} to \refeq{eq:proc3}. The emission of an additional photon leads to
the processes
\begin{eqnarray}
\label{eq:bremsproc1}
& \Pu_i \, \, \Pdbar_j & \to \Plp \Pnl \, \Pg \, \gamma \, ,\\
\label{eq:bremsproc2}
& \Pu_i \, \, \Pg      & \to \Plp \Pnl \, \Pd_j \, \gamma \, ,\\
\label{eq:bremsproc3}
& \Pdbar_j \, \, \Pg   & \to \Plp \Pnl \, \Pubar_i \, \gamma \, .
\end{eqnarray}
The Feynman diagrams contributing to the process \refeq{eq:bremsproc2} 
are shown in \reffi{fi:EW_real_udWg}.
Due to the emission of soft photons the real corrections
include soft singularities which are cancelled by the virtual
corrections independently of the details of the event selection or
recombination procedure. If the photon and the charged
leptons/quarks are recombined into a pseudo-particle (mimicking the
start of hadronic or electromagnetic showers) to form IR-safe
observables, all the remaining singularities arising from collinear
photon emission in the final state also cancel against the
corresponding singularities in the virtual corrections. This
requires that all the selection  cuts for a given observable are
blind to the distribution of momenta in collinear lepton--photon
configurations. The left-over collinear singularities due to
collinear photon emission off the initial-state quarks are absorbed
by a redefinition  of the PDFs. Technically, we use the dipole
subtraction formalism as specified for photon emission in  
\citeres{Dittmaier:1999mb,Dittmaier:2008md}
to isolate all the divergences  and observe the numerical
cancellation. Note that the MRSTQED2004 PDFs, which properly
account for all QED effects, are defined in the  DIS scheme with
respect to QED corrections, as explained in \citere{Diener:2005me}.

\bfi[t]
\begin{center}
\includegraphics{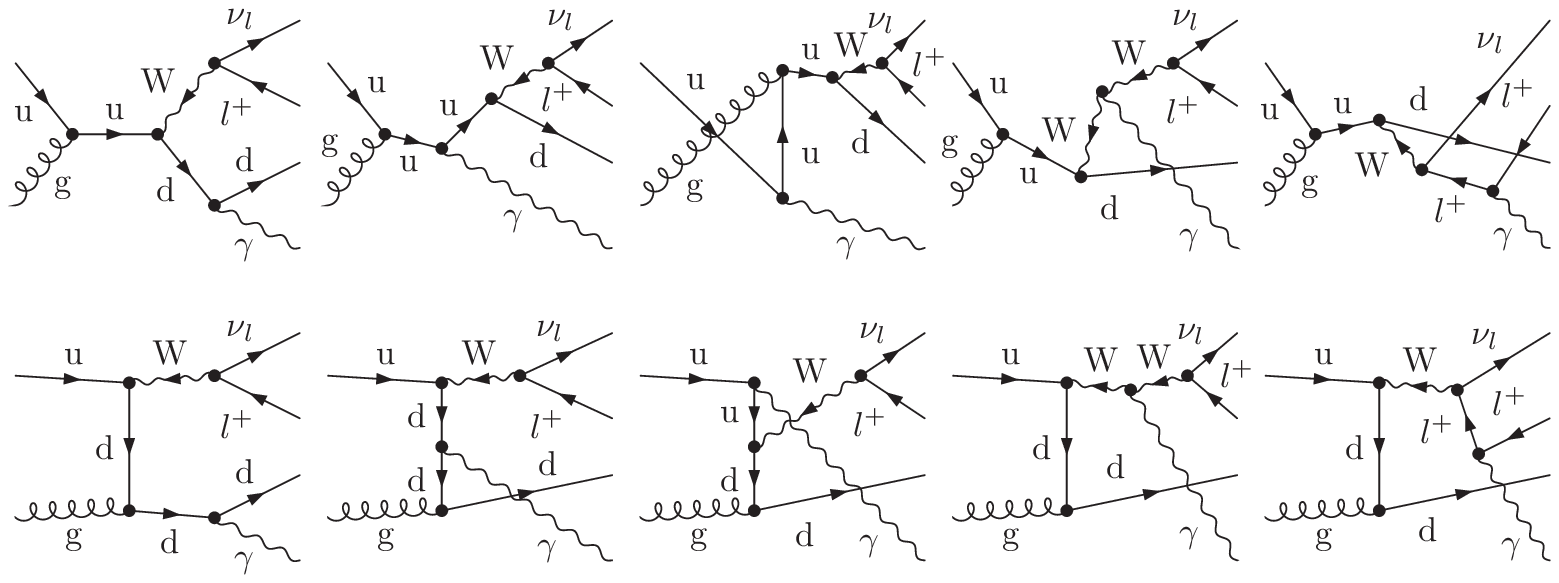}
\end{center}
\mycaption{\label{fi:EW_real_udWg}
Real photonic bremsstrahlung corrections to the LO process \refeq{eq:proc2}.}
\efi

For muons in the final state it is, however, experimentally possible
to separate collinear photons from the lepton, i.e.\ to observe
so-called ``bare'' muons. Hence, the resulting cross sections are not
collinear safe (i.e.\ the KLN theorem \cite{Kinoshita:1962ur} does not
apply), and the corresponding collinear singularities show up as
logarithms of the small lepton (muon) mass. The lepton mass cuts off
the collinear divergence in a physically meaningful way. 

In this work, we employ the extension \cite{Dittmaier:2008md} of the 
subtraction formalism \cite{Dittmaier:1999mb}, which allows one 
to calculate cross sections for
bare leptons, i.e.\ cross sections defined without any photon
recombination. Like in the standard subtraction formalism, it is
sufficient to calculate the real-emission matrix elements for the partonic
processes in the massless-fermion approximation. The main difference
between the subtraction variants of \citeres{Dittmaier:1999mb} and 
\cite{Dittmaier:2008md} concerns the
implementation of phase-space cuts. In the standard subtraction
formalism \cite{Dittmaier:1999mb} it is always assumed that 
the complete momentum of
the lepton and a collinear photon is subject to cuts (as it would be
the case after recombination), while the generalization
\cite{Dittmaier:2008md} allows
for non-collinear-safe cuts that resolve the distribution of the
momenta in collinear photon--lepton configurations. This more
general cut procedure in the non-collinear-safe case has to be
carefully implemented both in the real emission part and the
corresponding subtraction terms, in order to ensure the numerical
cancellation of singularities. Of course, the treatment of
non-collinear-safe cuts leads to a modification of the readded
subtraction part as well. In the formulation of \citere{Dittmaier:2008md} 
this modification assumes the form of an additional (+)-distribution
which contains the surviving mass singularity. This (+)-distribution
integrates to zero for collinear-safe observables so that the
formalism reduces to the well-known standard subtraction formalism.
For non-collinear-safe observables the additional logarithms of the
lepton mass in the final result are, thus, isolated analytically.
For a complete and detailed description of this more general
subtraction formalism we refer the reader to \citere{Dittmaier:2008md}. 
In the following we briefly outline how the logarithms of the lepton
mass for our specific example of bare muons in $\PW+\mathrm{jet}$
can be extracted. We have also used the two-cutoff phase-space
slicing to check the results for the EW corrections.

The logarithmically enhanced terms can be directly extracted from the readded
integrated subtraction terms. For a final-state emitter and a
final-state spectator they can be cast into the form [see (2.14) in
\citere{Dittmaier:2008md}]
\beqar
\int\rd\Phi_1\,|\M_{\sub,ij}(\Phi_1)|^2
&=& -\frac{\alpha}{2\pi} Q_i\sigma_i Q_j\sigma_j  \,
\int\rd\tilde\Phi_{0,ij}\,
\int_0^1 \rd z\, 
\nn\\[.3em]
&& {}\times
\left\{ \Gsub_{ij}(P_{ij}^2) \de(1-z) 
+ \left[\bcGsub_{ij}(P_{ij}^2,z)\right]_+ \right\}
\nn\\[.3em]
&& {}\times
|\M_0(\tilde p_i,\tilde p_j)|^2 \,
\Theta_{\cut}\Bigl(p_i=z\tilde p_i,k=(1-z)\tilde p_i,\tilde p_j,\{k_n\}\Bigr),
\label{eq:zintij}
\eeqar
where $Q_{i,j}$ are the fermion charges of emitter $i$ and spectator
$j$, $\tilde p_{i,j}$ the momenta, and $\si_{i,j}=\pm1$ correspond to the
charge flow [$\sigma_{i,j}=+1 (-1)$ for incoming (outgoing) fermions
and outgoing (incoming) antifermions]. We denote the LO matrix element
by $\M_0(\tilde p_i,\tilde p_j)$, and $\rd\tilde\Phi_{0,ij}$ indicates
the integration over the LO phase space.  The cuts on the phase space
are implemented via the theta function $\Theta_{\cut}$ which is zero
if a momentum configuration does not pass the cut and one otherwise.
Note that $\Theta_{\cut}$ is a function of $p_i=z\tilde p_i$, the
photon momentum $k=(1-z)\tilde p_i$, and all other momenta, i.e.\ the
momenta of the emitter and the photon after emission are reconstructed
from the recombined momentum $\tilde p_i$. Here, collinear safety is
not guaranteed. The result is indeed presented in a form, where the endpoint
contribution $\Gsub_{ij}$, known from the collinear-safe subtraction
formalism~\cite{Dittmaier:1999mb}, 
is explicitly extracted. The term $\Gsub_{ij}$ contains all
the fermion and photon mass logarithms that are cancelled by the
virtual corrections in mass regularization. The delta-function
$\de(1-z)$ guarantees that this result is unchanged no matter if the
cuts are collinear safe or not. As mentioned above, 
the additional contribution
$\bcGsub_{ij}$ integrates trivially to zero if the cuts do not depend
on $z$ due to the (+)-distribution with respect to the $z$
integration. However, there is an additional contribution for
non-collinear-safe cuts. For $\bcGsub_{ij}$, one finds~\cite{Dittmaier:2008md}
\beqar
\bcGsub_{ij}(P_{ij}^2,z) &=&
P_{ff}(z)\,\biggl[ \ln\biggl(\frac{P_{ij}^2 z}{m_i^2}\biggr)-1\biggl]
+(1+z)\ln(1-z) +1-z, 
\label{eq:bcGsubij2}
\eeqar
where $P^2_{ij}=(\tilde p_i + \tilde p_j)^2$ and the splitting function reads
\beq
P_{ff}(z) = \frac{1+z^2}{1-z}.
\eeq
This contribution contains the logarithmic contributions we are after.
Omitting non-logarithmic terms we find for  a particular emitter--spectator 
contribution
\beqar
\label{eq:non_coll_res}
&& \int\rd\Phi_1\, |\M_{\sub,ij}(\Phi_1)|^2
= \frac{\alpha}{2\pi}  \,
\int\rd\tilde\Phi_{0,ij}\,
\int_0^1 \rd z\, \Gamma_{ij}(z)
|\M_0(\tilde p_i,\tilde p_j)|^2 \,
\nn\\[.3em]
&& \quad\quad\quad {}\times
\left[\Theta_{\cut}\Bigl(p_i=z\tilde p_i,k=(1-z)\tilde p_i,\tilde p_j,\{k_n\}\Bigr) -
\Theta_{\cut}\Bigl(p_i=\tilde p_i,k=0,\tilde p_j,\{k_n\}\Bigr)\right]
\nn\\[.5em]
&& \quad\quad\quad {} + \mbox{non-singular terms},
\eeqar
where the (+)-distribution has been made explicit and we have defined the
function
\beq
\Gamma_{ij}(z) = - Q_i\sigma_i Q_j\sigma_j P_{ff}(z)\,
                 \ln\biggl(\frac{P_{ij}^2 z}{m_i^2}\biggr) \,.
\eeq
For an initial-state spectator $a$, the contribution looks exactly the same, 
apart from the replacement $P_{ij}^2 \to -P_{ia}^2$. Using charge conservation
($\sum_j\si_j Q_j + \sum_a\si_a Q_a=-\si_i Q_i$),
one finds for the sum of all subtraction terms corresponding to the emission by
a final-state fermion
\beq
\label{eq:Gamma}
\Gamma_{i}(z) = \sum_{k} \Gamma_{ik}(z)
= Q_i^2 \,
P_{ff}(z)\, 
\ln\biggl(\frac{Q^2}{m_i^2}\biggr) -
\sum_{k}
Q_i\sigma_i Q_k\sigma_k P_{ff}(z)
\ln\biggl(\frac{|P_{ik}^2| z}{Q^2}\biggr) \, ,
\eeq
where $k$ runs over all initial-state and final-state spectators, and
we have introduced the scale $Q$ to combine logarithms of the fermion
masses. The first term is exactly the first-order leading-logarithmic
result in the structure-function approach to final-state
radiation~\cite{Kuraev:1985hb} (see also \citere{Brensing:2007qm} for
a specific application to \PW-boson production). In the
structure-function approach, $Q$ is interpreted as a factorization
scale. Evidently, in our complete $\Oa$ calculation this scale
dependence is absent.
Moreover, for real emission events, the full kinematics of the
lepton--photon splitting is contained in the calculation, i.e.\ we are
not restricted to the strict collinear limit. For integrated
$\PW+\mathrm{jet}$ cross sections, the above result implies an
additional negative correction, since we demand a minimum transverse
momentum for the charged leptons.  Hence, the cut condition in the
first theta-function of \refeq{eq:non_coll_res} is more restrictive.
For distributions, logarithmically enhanced corrections that would
contribute to a given transverse-momentum bin with recombination are
now shifted to a bin corresponding to $z \tilde p_i$ which has, of
course, less $p_\rT$ than $\tilde p_i$.

Photons and QCD partons always have to be recombined into a single jet
if they are sufficiently collinear. This leads to collinear-safe
observables if the selection cuts respect the recombination
procedure. However, the recombination induces a problem for subprocess
\refeq{eq:bremsproc1}. If the photon and the gluon are accidentally
collinear (of course there is no collinear enhancement for these
configurations) arbitrarily soft gluons can still pass
the jet selection due to a collinear photon. There is a soft-gluon
divergence induced by this simple recombination procedure that would
be cancelled by the virtual QCD corrections to $\PW+\mathrm{photon}$ production,
as e.g.\ worked out in \citere{Hollik:2007sq}. To avoid the singularity,
one has to distinguish $\PW+\mathrm{photon}$ and $\PW+\mathrm{jet}$ production by means of a
more precise event definition employing a cut on the maximal energy or
transverse momentum fraction of a photon inside a given jet. However,
this procedure spoils the collinear safety of the event definition in
subprocesses \refeq{eq:bremsproc2} and \refeq{eq:bremsproc3}. Using
again the subtraction formalism~\cite{Dittmaier:2008md} to extract the
problematic collinear terms, the appearance of an unphysical quark-mass 
logarithm in the final result signals the necessity to include
non-perturbative physics to properly describe the emission of a photon
by a quark for exclusive final states. Using dimensional
regularisation, the quark-mass logarithm translates into a collinear
$1/\epsilon$ pole in the final results, where $\epsilon$ quantifies
the deviation from the four-dimensionality of space--time. Fortunately,
the non-perturbative contribution to the class of events we want to
exclude has been measured at LEP in photon+jet events. In these events
a photon carries almost all the energy of a radiating quark 
in a hadronic \PZ-boson decay. The relevant collinear
physics can be factorized from the underlying hard process and can be
cast into a process-independent quark-to-photon fragmentation function
$D_{q\to\gamma}(z_{\ga})$, where $z_{\ga}$ is the energy fraction of the
photon in the collinear quark--photon configuration.

In analogy to the absorption of initial-state collinear singularities
into PDFs, the perturbative singularity can be absorbed into an NLO
definition of the fragmentation function~\cite{Glover:1993xc}. 
However, in contrast to the PDF analogue, the LO fragmentation
function vanishes. 
Using dimensional regularization (DR) and the \MSbar\ factorization
scheme, one finds
\beq
\label{eq:msbar}
D^{\mathrm{DR}}_{q\to\gamma}(z_{\gamma}) = 
\frac{\alpha Q_q^2}{2 \pi}
P_{q\to\gamma}(z_{\gamma}) 
\left( \frac{(4\pi)^\epsilon}{\Gamma(1-\epsilon)}\,
\frac{1}{\epsilon}
+\ln \frac{\mu^2}{\mu_{\mathrm{F}}^2} \right)
+ D^{\mathrm{ALEPH},\overline{\mathrm{MS}}}_{q\to\gamma}(z_{\ga},\mu_{\mathrm{F}}) ,
\eeq 
where $D=4-2\epsilon$ is the dimension of space--time, $\mu$ is the
arbitrary reference mass of dimensional regularization, $\mu_{\mathrm{F}}$ 
is the factorization scale for the fragmentation process, and
the quark-to-photon splitting function is given by 
\beq
P_{q\to\gamma}(z_{\ga})=\frac{1+(1-z_{\ga})^2}{z_{\ga}} .
\eeq 
Translating the \MSbar\ definition
into the mass regularization (MR) scheme we find
\beq
\label{eq:frag}
D^{\mathrm{MR}}_{q\to\gamma}(z_{\gamma})
 =  \frac{\alpha Q_q^2}{2 \pi}
P_{q\to\gamma}(z_{\gamma}) \left(\ln \frac{m_q^2}{\mu_{\mathrm{F}}^2} + 2 \ln
z_{\ga} + 1 \right) + D^{\mathrm{ALEPH},\overline{\mathrm{MS}}}_{q\to\gamma}(z_{\ga},\mu_{\mathrm{F}}) .
\eeq 
As indicated by the superscript, we will employ
the parametrization of the fragmentation function
used by the ALEPH collaboration to fit the data~\cite{Buskulic:1995au},
\beq 
D^{\mathrm{ALEPH},\overline{\mathrm{MS}}}_{q\to\gamma}(z_{\ga},\mu_{\mathrm{F}}) = \frac{\alpha
  Q_q^2}{2 \pi} \left( P_{q\to\gamma}(z_{\ga}) \ln
\frac{\mu_{\mathrm{F}}^2}{(1-z_{\ga})^2 \mu_0^2} + C \right) .  
\eeq 
Here, the
constants $\mu_0^2$ and $C$ are fit parameters and the dependence of
the complete fragmentation function on the factorization scale $\mu_{\mathrm{F}}$
cancels by construction. We use the result of a one-parameter fit
where $C$ is constraint to $C=-1-\ln(\MZ^2/(2 \mu_0^2))$ resulting in
\beq \mu_0 = 0.14 \, \mathrm{GeV} \quad \mathrm{and} \quad C = -13.26
\, . \eeq 
Note that we are interested in the fragmentation function
to subtract the non-perturbative part of the perturbatively
well-defined inclusive (collinear-safe) cross section in which the
photon-energy fraction $z_{\ga}$ becomes large\footnote{
In contrast, if one did not recombine photons and partons at all to avoid 
the soft-gluon pole the result would be sensitive to non-perturbative collinear 
physics, more precisely to the fragmentation function in the whole range 
of $z_{\ga}$, in particular for small $z_{\ga}$ where the fragmentation 
function has not been measured.}.
Hence, we are not
sensitive to the soft-photon pole of the splitting function for
$z_{\ga}\to 0$. We convolute the fragmentation function \refeq{eq:frag}
with the LO cross section if $z_{\ga}$ is larger than a
given cutoff. The result is subtracted from the inclusive result
together with the perturbative parts, captured by the subtraction
formalism, and the unphysical dependence of the cross section on the quark
mass disappears.

The real corrections due to NLO QCD are less subtle. Additional gluon emission 
and application of crossing symmetry leads to the processes 
\begin{eqnarray}
\label{eq:bremsproc4}
& \Pu_i \, \, \Pdbar_j & \to \Plp \Pnl \, \Pg \, \Pg \, ,\\
\label{eq:bremsproc5}
& \Pu_i \, \, \Pg      & \to \Plp \Pnl \, \Pd_j \, \Pg \, ,\\
\label{eq:bremsproc6}
& \Pdbar_j \, \, \Pg   & \to \Plp \Pnl \, \Pubar_i \, \Pg \, , \\
\label{eq:bremsproc7}
& \Pg\, \, \Pg   & \to \Plp \Pnl \, \Pubar_i \, \Pd_j \, ,
\end{eqnarray}
and the external gluon present at LO may also split into two quarks, 
inducing the processes
\begin{eqnarray}
\label{eq:bremsproc8}
& \Pu_i \, \, \Pdbar_j & \to \Plp \Pnl \, q_k \, \bar q_k \, ,\\
\label{eq:bremsproc9}
& \bar q_k \, \, \Pdbar_j & \to \Plp \Pnl \, \Pubar_i \, \bar q_k \, ,\\
\label{eq:bremsproc10}
& q_k \, \, \Pdbar_j & \to \Plp \Pnl \, q_k \, \Pubar_i 
\quad (q_k\ne\Pu_i, q_k\ne\Pd_j) \, ,\\
\label{eq:bremsproc11}
& \Pu_i \, \, \bar q_k & \to \Plp \Pnl \, \Pd_j \, \bar q_k 
\quad (\bar q_k\ne\Pdbar_j, \bar q_k\ne\Pubar_i) \, ,\\
\label{eq:bremsproc12}
& \Pu_i \, \, q_k & \to \Plp \Pnl \, q_k \, \Pd_j \, ,\\
\label{eq:bremsproc13}
& q_k \, \, \bar q_k & \to \Plp \Pnl \, \Pubar_i \, \Pd_j \, ,
\end{eqnarray}
where $q_k$ stands for up-type quarks $\Pu_k$ with $k=1,2$ or for
down-type quarks $\Pd_k$ with $k=1,2,3$.  Note that the Feynman
diagrams are different in the two cases $i=k$ ($j=k$) and $i\ne k$
($j\ne k$). Taking this difference into account, the remaining sums
over flavour can again efficiently be performed when convoluting the
squared matrix elements with PDFs. As in the EW case, we use the
dipole subtraction method \cite{Catani:1996vz} to extract the IR
singularities analytically from the numerical phase-space integration.
Absorbing all the collinear singularities due to initial-state
splittings into the relevant PDFs, the remaining collinear and soft
divergences cancel all the divergences of the one-loop QCD corrections
for processes \refeq{eq:proc1}--\refeq{eq:proc3}.  Here, also the
bottom-quark PDF enters the NLO prediction. For example a bottom quark
from a proton can emit a gluon which subsequently takes part in the
hard process.

Turning to the photon-induced processes, the corresponding bremsstrahlung 
processes are
\begin{eqnarray}
\label{eq:bremsproc14}
& \Pu_i \, \, \ga      & \to \Plp \Pnl \, \Pd_j \, \Pg \, ,\\
\label{eq:bremsproc15}
& \Pdbar_j \, \, \ga   & \to \Plp \Pnl \, \Pubar_i \, \Pg \, , \\
\label{eq:bremsproc16}
& \Pg \, \, \ga   & \to \Plp \Pnl \, \Pubar_i \, \Pd_j \, .
\end{eqnarray}
All singularities cancel those in the virtual NLO QCD
corrections or are absorbed into PDFs.

There is yet another class of corrections contributing at $\mathcal{O}(\alpha^3
\alpha_{\mathrm{s}})$. For the six-fermion processes 
\bfi[fp]
\includegraphics{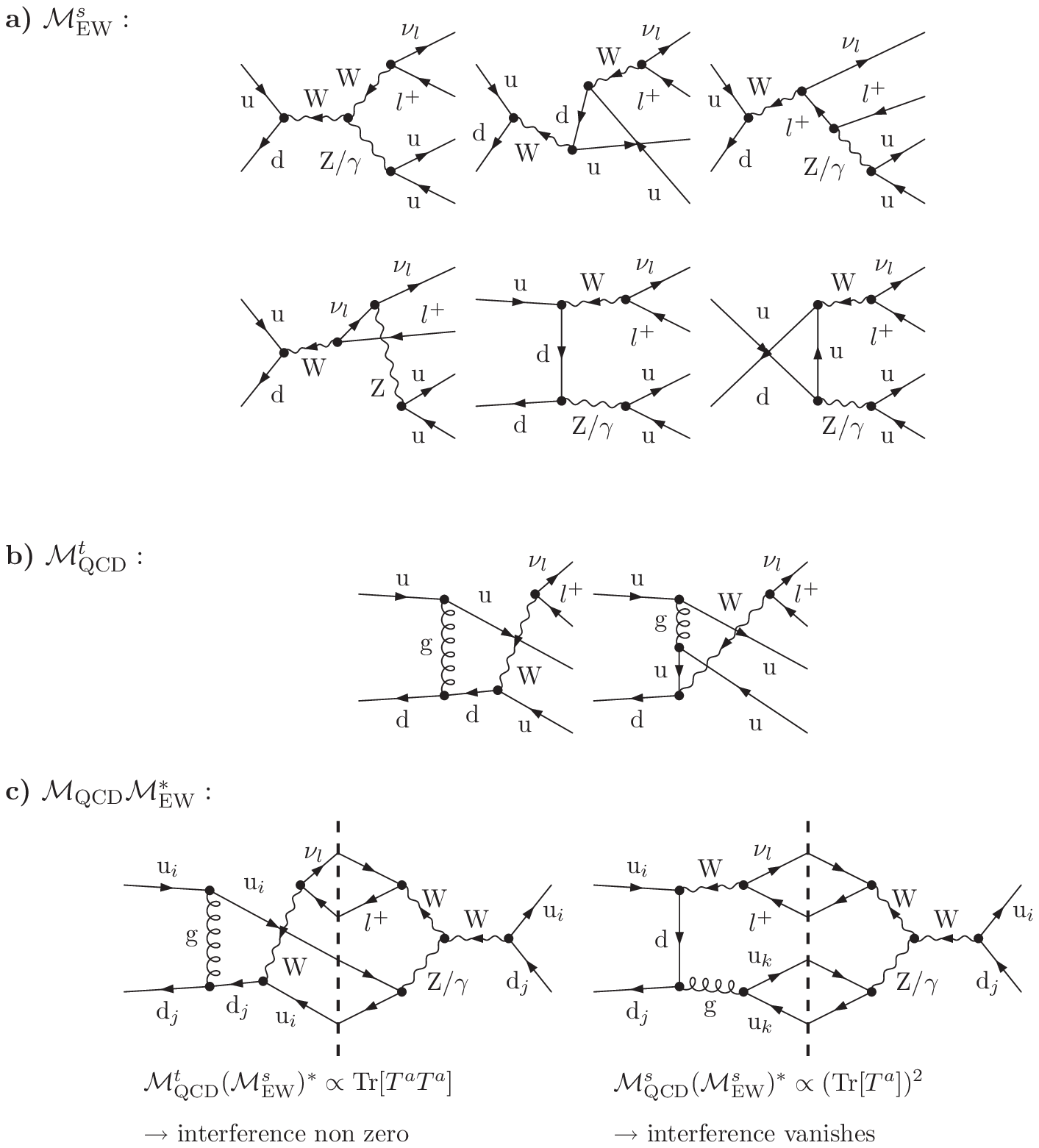}
\mycaption{\label{fi:EW_Interf_udWg} The interference between EW and QCD
  diagrams:
  a) EW diagrams with $s$--channel-like colour flow
  for process \refeq{eq:bremsproc8} with $q_k=\Pu_k$. b) QCD diagrams
  with $t$--channel-like colour flow.  c) EW and QCD diagrams 
  of both $s$- or $t$-type do not contribute (right).  However, for $i =
  k$, there is an interference contribution from diagrams of
  different types that is non-zero. The full interference term
  for the partonic process \refeq{eq:bremsproc8} reads $
  2\,\mathrm{Re} [\M^t_{\mathrm{QCD}} ( \M_{\mathrm{EW}}^s)^* +
  \M^s_{\mathrm{QCD}} (\M_{\mathrm{EW}}^t)^*]$. All other partonic
  interference contributions with the same flavour structure can be
  obtained by applying the same crossing procedure to
  $\M^{s/t}_{\mathrm{QCD}}$ and $\M^{s/t}_{\mathrm{EW}}$.}
\efi
\refeq{eq:bremsproc8}--\refeq{eq:bremsproc13}
with two identical quarks, diagrams with gluon exchange
can interfere with purely EW diagrams. Exemplarily, the relevant diagrams for
one of the contributing subprocesses are shown in \reffi{fi:EW_Interf_udWg}. 
The result is non-singular due to the restrictions from colour flow, 
but only if all fermions are distinct the
interference contribution vanishes. We have also included these corrections
in our calculation. However, their effect turns out to be 
phenomenologically negligible. Diagrams with an
internal top propagator and two \PW\ bosons do not contribute if mixing with
quarks of the third generation is neglected.

\section{Numerical results}
\label{se:numres}

\subsection{Input parameters and setup}
\label{se:SMinput}

The relevant SM input parameters are
\begin{equation}\arraycolsep 2pt
\begin{array}[b]{lcllcllcl}
\GF & = & 1.16637 \times 10^{-5} \GeV^{-2}, \quad&
\Lambda_\mathrm{QCD} &=& 239\MeV,&
\alpha_{\mathrm{s}}(\MZ) &=& 0.11899 , 
\\
\MW^{\OS} & = & 80.398\GeV, &
\Gamma_\PW^{\OS} & = & 2.141\GeV, \\
\MZ^{\OS} & = & 91.1876\GeV, &
\Gamma_\PZ^{\OS} & = & 2.4952\GeV, &
M_\PH & = & 120\GeV, \\
m_\Pe & = & 0.510998910\MeV, &
m_\mu &=& 105.658367\MeV,\quad &
m_\Pt & = & 172.6\;\GeV,
\\
|V_{\Pu\Pd}| & = & |V_{\Pc\Ps}| = 0.974, &
|V_{\Pu\Ps}| & = & |V_{\Pc\Pd}| = \sqrt{1 - |V_{\Pc\Ps}|^2}, 
\end{array}
\label{eq:SMpar}
\end{equation}
which essentially follow \citere{Amsler:2008zzb}. The CKM
matrix is included via global factors in the partonic cross sections
for the different possible quark flavours. Within loops the CKM matrix is 
set to unity, because its effect is negligible there.

Using the complex-mass scheme~\cite{Denner:2005fg}, we employ a fixed
width in the resonant W- and Z-boson propagators in contrast to the
approach used at LEP and Tevatron
to fit the W~and Z~resonances, where running
widths are taken. Therefore, we have to convert the ``on-shell'' (OS)
values of $M_V^{\OS}$ and $\Ga_V^{\OS}$ ($V=\PW,\PZ$), resulting
from LEP and Tevatron, to the ``pole values'' denoted by $M_V$ and $\Ga_V$. The
relation between the two sets of values is given by
\cite{Bardin:1988xt}
\beq\label{eq:m_ga_pole}
M_V = M_V^{\OS}/
\sqrt{1+(\Ga_V^{\OS}/M_V^{\OS})^2},
\qquad
\Ga_V = \Ga_V^{\OS}/
\sqrt{1+(\Ga_V^{\OS}/M_V^{\OS})^2},
\eeq
leading to
\beqar
\begin{array}[b]{r@{\,}l@{\qquad}r@{\,}l}
\MW &= 80.370\ldots\GeV, & \GW &= 2.1402\ldots\GeV, \\
\MZ &= 91.153\ldots\GeV,& \GZ &= 2.4943\ldots\GeV.
\label{eq:m_ga_pole_num}
\end{array}
\eeqar
We make use of these mass and width parameters in the numerics discussed below,
although the difference between using $M_V$ or $M_V^{\OS}$ would be
hardly visible.

As explained in \refse{se:setup}, we adopt the $\GF$ scheme, where the
electromagnetic coupling $\alpha$ is set to $\alpha_{\GF}$. 
In this scheme the electric-charge renormalization constant does not contain
logarithms of the light-fermion masses, in contrast to the $\alpha(0)$
scheme, so that the results become practically
independent of the light-quark masses.

The ${\cal O}(\alpha)$-improved MRSTQED2004 set of
PDFs~\cite{Martin:2004dh} is used throughout implying the value
of $\alpha_{\mathrm{s}}(\MZ)$ stated in \refeq{eq:SMpar}. We use 
standard two-loop running of the strong coupling constant in the 5-flavour
scheme with $\Lambda_\mathrm{QCD} = 239\,\mathrm{MeV}$. Since 
the MRSTQED2004 PDF set has been released, there have been 
considerable improvements for PDFs, in particular with respect to
the heavy-flavour treatment. Since recent PDF sets do not include
QED effects we stick to MRSTQED2004 for theoretical consistency.
Hence, all the absolute values for cross sections lack the
recent PDF improvements. However, the presented relative corrections
should be more stable with respect to variations in the PDFs than
absolute predictions.

The QCD and QED factorization scales as well as the renormalization
scale are always identified.  For low-$p_{\rT}$ jets, the scale of the
process is given by the invariant mass of the leptons which in turn
peaks around \MW\ for resonant \PW-boson production.  Hence, one
natural choice is the \PW-boson mass, i.e.\ 
$\mu_\mathrm{R}=\mu_\mathrm{F}=\MW$.  For high-$p_{\rT}$ jets, well
beyond the \PW-boson scale, however, the relevant scale is certainly
larger, and the QCD emission from the initial state is best modelled
by the $p_{\rT}$ of the jet itself (see e.g.\ \citere{Bauer:2009km}). 
To interpolate between the two regimes, we alternatively use
\begin{equation}
\label{eq:scale_choice}
\mu=\mu_\mathrm{R}=\mu_\mathrm{F}=\sqrt{\MW^2+(p_{\rT}^\mathrm{had})^2} \, , 
\end{equation}
where $p_{\rT}^\mathrm{had}$ is given by the $p_{\rT}$ of the summed
four-momenta of all partons, i.e.\ quarks and/or gluons in the final
state. At LO, $p_{\rT}^\mathrm{had}$ is simply the $p_{\rT}$ of the
one final-state jet. We present results for both scale choices.

\subsection{Phase-space cuts and event selection}
\label{se:cuts}

In order to define IR-safe observables for the process
$\Pp\Pp/\Pp\bar\Pp\to\PW^+ + \mathrm{jet} \to \Pl^+\nu_\Pl + \mathrm{jet} + \X$ 
we recombine final-state partons and photons to pseudo-particles
and impose a set of 
phase-space cuts as detailed in the following subsections.

\subsubsection{Recombination}

To define the recombination procedure and the separation cuts, we use
the variables $R_{ij} = \sqrt{(y_{i}-y_{j})^2+\phi_{ij}^2}$, where
$y_{i}$ denotes the rapidity $y = \frac{1}{2} \ln [(E + p_\RL)/(E -
p_\RL)]$ 
of particle $i$ and $\phi_{ij}$ is the azimuthal angle in the
transverse plane between the particles $i$ and $j$. In the definition
of the rapidity, $E$ denotes the particle's energy and $p_{\RL}$ the
momentum along the beam axis. The recombination procedure, where we
simply add four-momenta to form a pseudo-particle, works as follows:
\begin{enumerate}
\item
For observables with bare muons we do not recombine photons and leptons. For inclusive
observables, a photon and a lepton are recombined for $R_{\gamma l} < 0.1$.
\item A photon and a parton $a$ (quark or gluon) are recombined for
$R_{\gamma a} < 0.5$. In this case, we use the energy		      
fraction of the photon inside the jet, $z_{\gamma}=		      
E_{\gamma}/(E_{\gamma}+E_a)$, 
to distinguish between     
$\mathrm{W}+\mathrm{jet}$ and $\mathrm{W}+\gamma$ production. If      
$z_{\gamma} > 0.7$, the event is regarded as a part of  	      
$\mathrm{W}+\gamma$ production and rejected because it lacks any      
other hard jet at NLO. This event definition is not collinear safe    
and requires the use of quark-to-photon fragmentation functions to  
include the non-perturbative part of the quark--photon splitting as   
explained in \refse{se:real}.					      
Our results are not very sensitive to the specific choice of the cut
on $z_{\gamma}$.
\item
Two partons $a,b$ are recombined for $R_{ab} < 0.5$. For our simple 
final-state configurations, this procedure is equivalent to the Tevatron Run II 
$k_{\rT}$-algorithm \cite{Blazey:2000qt} for jet reconstruction with resolution parameter $D=0.5$.
\end{enumerate}

Technically, we perform a possible photon--lepton recombination before the
photon--parton recombination. This procedure is IR safe because the
triple-soft/collinear situation that a photon should have been first recombined 
with a parton, but was erroneously first recombined with a lepton,
is excluded by our basic cuts.

\subsubsection{Basic cuts}
\label{se:basic_cuts}

After applying the recombination procedure of the previous section we define
$\PW+\mathrm{jet}$ events by the following basic cuts:
\begin{enumerate}
\item
A partonic object (after a possible
recombination) is called a jet if its transverse momentum 
$p_{\rT}$ is larger than $p^{\mathrm{cut}}_{\rT,\mathrm{jet}} = 25$ GeV. 
Events are required to include at least one jet.
\item
We demand a charged lepton with transverse momentum 
$p_{\rT,\Pl} > 25\; \mathrm{GeV}$ and a missing momentum
$\dsl{p}_{\rT} > 25\; \mathrm{GeV}$.
\item
The events have to be central, i.e.\ the lepton and at least one jet have 
to be produced in the rapidity range $|y| < y_{\mathrm{max}} = 2.5$. 
\item
The lepton has to be isolated, i.e.\ the event is discarded if the distance
between the lepton and a jet $R_{l\mathrm{jet}}$ is smaller than 0.5.\\ 
The lepton--jet separation is also required for jets with $|y| > y_{\mathrm{max}}$.
It is important to exclude low-$p_{\rT}$ partons from the
lepton--jet separation procedure (guaranteed by step 1.),
since otherwise observables would not be IR safe. 
\end{enumerate} 

While the EW corrections differ for final-state electrons and
muons without photon recombination, the corrections become universal
in the presence of photon recombination, since the lepton-mass
logarithms cancel in this case, in accordance with the KLN theorem.
Numerical results are presented for photon recombination and for bare
muons.

For certain observables, we apply a jet veto against a second hard jet. 
To be specific, we veto any sub-leading jet with $p_{\rT} >
p_{\rT,\Pj_1}/2$, where $p_{\rT,\Pj_1}$ denotes the
$p_{\rT}$ of the ``leading'' jet, i.e.\ the one with maximal $p_{\rT}$.

\subsection{Results on cross sections}
\label{se:CSresults}

We first consider $\PW^+$~production in association with a jet at the
LHC, i.e.\ a \Pp\Pp~initial state with a centre-of-mass (CM) energy of
$\sqrt{s}=14\TeV$.

We present the LO cross section $\sigma_0$ and various types of corrections $\de$,
defined relative to the LO cross section by $\sigma =
\sigma_0\times\left(1+\de\right)$. Concerning the EW corrections, we distinguish the
cross section $\sigma^{\mu^+ \nu_\mu}_\mathrm{EW}$ for bare muons and
$\sigma^{\mathrm{rec}}_\mathrm{EW}$ for which lepton--photon recombination is employed as
defined above. Accordingly, the corresponding corrections are labelled $\delta^{\mu^+
\nu_\mu}_\mathrm{EW}$ and $\delta^\mathrm{rec}_\mathrm{EW}$, respectively. An additional label
specifies which renormalization and factorization scale is used. Either we use the
fixed scale ($\mu=\MW$) or we determine the scale on an event-by-event basis by the
kinematical configuration of the final state (var), as specified in \refeq{eq:scale_choice}. 
For the EW corrections the difference is not expected to be large, 
since the LO and the NLO results depend on the renormalization scale 
for $\alpha_{\mathrm{s}}$ and the QCD factorization scale
in the same way. However, for the QCD part a sensible scale choice
can be crucial for the stability of the perturbative series. 
Accordingly, the QCD corrections are labelled $\delta_\mathrm{QCD}^{\mu=\MW}$
for a fixed scale choice and $\delta_\mathrm{QCD}^\mathrm{var}$ for the scale
choice defined in \refeq{eq:scale_choice}.

As shown below, the QCD corrections become larger and larger with
increasing $p_{\rT}$ of the leading jet.
The increase in the cross section results from a new kinematical
configuration which is available for the $\PW+2\,\mathrm{jets}$ final state. 
The large $p_{\rT}$ of the leading jet is not balanced by the leptons, as
required at LO, but by the second jet. Hence, we encounter the
production of 2 jets where one of the quark lines radiates a
relatively soft \PW\ boson. This part of the cross section, which does
not really correspond to a true NLO correction to $\PW+\mathrm{jet}$
production, can be separated by employing a veto against a second hard
jet in real-emission events. Hence, we present NLO QCD corrections
with a jet veto, ($\delta^{\mu=\MW}_\mathrm{QCD,veto}$,
$\delta^\mathrm{var}_\mathrm{QCD,veto}$), and without a jet veto,
($\delta^{\mu=\MW}_\mathrm{QCD}$, $\delta^\mathrm{var}_\mathrm{QCD}$).

Using a jet veto based on a fixed $p_{\rT}$ value for the second jet
is not well suited. It will either cut away relatively collinear
emission events in the high-$p_{\rT}$ tails of the leading-jet
distribution (leading to large negative corrections) or it has to be
chosen too large to be effective in the intermediate-$p_{\rT}$\ parts of the
distribution. Hence, we veto any sub-leading jet with $p_{\rT} >
p_{\rT,\Pj_1}/2$, where $p_{\rT,\Pj_1}$ denotes the
$p_{\rT}$ of the leading jet.
As shown below, 
this jet veto indeed effectively
removes events with back-to-back kinematics.

We also investigate the impact of the photon-induced tree-level
processes \refeq{eq:proc4} and \refeq{eq:proc5} and the corresponding
NLO QCD corrections including the real-emission processes
\refeq{eq:bremsproc14}, \refeq{eq:bremsproc15}, and
\refeq{eq:bremsproc16}.  Since even the LO photon-induced
cross section is a small
effect, we show its relative impact $\delta_{\gamma,\mathrm{Born}}$
with respect to the LO cross section where initial states with photons
are not taken into account. Including the NLO QCD corrections, the
relative impact of the full NLO cross section is denoted by
$\delta_{\gamma,\mathrm{NLO}}$. The impact of the interference
contribution introduced at the end of \refse{se:real} is denoted by
$\delta_{\Int}$. Additional labels again indicate the scale choice and
the usage of a jet veto.

\refta{ta:ptl_LHC} shows the LO predictions and the above
corrections for different cuts on the $p_{\rT}$ of the charged lepton
$p_{\rT,\Pl}$. All other cuts and the corresponding event selection
follow our default choice 
as introduced in \refse{se:cuts}. All
integrated cross sections and, hence, the corrections are dominated by
events close to the lowest accepted $p_{\rT,\Pl}$, as can be seen by
the rapid decrease of the integrated cross section when increasing the
$p_{\rT,\Pl}$ cut.

\begin{table}
 																	$$ \begin{array}{c|rrrrrr}
 								    \multicolumn{7}{c}{\Pp\Pp \to \Plp \Pnl\; \mathrm{jet} + X \;\mbox{at} \;\sqrt{s} =14 \TeV} \\
 	      \hline p_{\rT,\Pl} / \GeV & 25-\infty \;\;\;\; & 50-\infty \;\;\;\; & 100-\infty \;\;\; & 200-\infty \;\;\; & 500-\infty \;\;\; & 1000-\infty \; \\ 
 																		      \hline\hline
 \si_{\born}^{\mu = \MW}/\fba			       \; & \; 508568(11)      \; & \; 163715(5)       \; & \; 13095.5(4)      \; & \; 1484.16(4)      \; & \; 44.476(1)       \; & \; 1.37894(5)      \\ 
 \si_{\born}^{\mathrm{var}}/\fba		       \; & \; 501826(11)      \; & \; 159482(5)       \; & \; 11481.8(4)      \; & \; 1124.67(3)      \; & \; 25.8791(9)      \; & \; 0.64346(2)      \\ 
    \hline \hline													  
 \de_{\EW}^{\mu^+\nu_{\mu}\,,\mathrm{var}}/\%	       \; & \; -3.0\phz        \; & \; -5.4\phz        \; & \; -9.0\phz        \; & \; -14.8\phz       \; & \; -25.8(1)        \; & \; -36.7(1)        \\ 
 \de_{\EW}^{\mathrm{rec}\,,\mathrm{var}}/\%	       \; & \; -2.2\phz        \; & \; -3.2\phz        \; & \; -6.5\phz        \; & \; -11.7\phz       \; & \; -21.6(1)        \; & \; -31.4(1)        \\ 
    \hline \hline													  
 \de_{\QCD}^{\mu = \MW}/\%			       \; & \; 48.2(1)         \; & \; 34.6(1)         \; & \; 50.9(1)         \; & \; 30.3(1)         \; & \; -15.9(1)        \; & \; -60.4(1)        \\ 
 \de_{\QCD}^{\mathrm{var}}/\%			       \; & \; 47.9(1)         \; & \; 34.1(1)         \; & \; 54.6(1)         \; & \; 46.2(1)         \; & \; 27.5(1)         \; & \; 6.1(1)	       \\ 
    \hline \hline													  
 \de_{\ga,\born}^{\mathrm{var}}/\%		       \; & \; 0.4\phz         \; & \; 0.6\phz         \; & \; 1.7\phz         \; & \; 2.7\phz         \; & \; 4.1(1)	       \; & \; 4.9(1)	       \\ 
   \hline														  
 \de_{\ga,\NLO}^{\mathrm{var}}/\%		       \; & \; 0.4\phz         \; & \; 0.6\phz         \; & \; 1.7\phz         \; & \; 2.6\phz         \; & \; 4.1(1)	       \; & \; 5.0(1)	       \\ 
    \hline \hline													  
 \de_{\Int}^{\mathrm{var}}/\%			       \; & \; 0.05\phantom{()}\; & \; 0.02\phantom{()}\; & \; 0.02\phantom{()}\; & \; 0.01\phantom{()}\; & \; -0.05\phantom{()}\; & \; -0.1\phz       \\ 
  \end{array} $$
\mycaption{\label{ta:ptl_LHC} Integrated cross sections for different
  cuts on the lepton transverse momentum at the LHC. We show the LO
  results for both a variable and a constant scale. The relative EW
  corrections $\de_{\EW}$ are given with and without lepton--photon
  recombination. The QCD corrections $\de_{\QCD}$ are presented for 
  a fixed as well as a variable scale. The corrections
  due to photon-induced processes $\de_{\ga}$, and the contributions
  from interference terms $\de_{\Int}$ are presented for a variable
  scale. The error from the Monte Carlo integration for the last digit(s) 
  is given in parenthesis as far as significant.
  See text for details.}
\end{table}

\begin{table}
 																	$$ \begin{array}{c|rrrrrr}
 								    \multicolumn{7}{c}{\Pp\Pp \to \Plp \Pnl\; \mathrm{jet} + X \;\mbox{at} \;\sqrt{s} =14 \TeV} \\
 		    \hline M_{\rT,\Pl\Pnl} / \GeV & 50-\infty \;\;\; & 100-\infty \;\; & 200-\infty \;\; & 500-\infty \;\; & 1000-\infty \; & 2000-\infty \;\; \\ 
 																		      \hline\hline
 \si_{\born}^{\mu = \MW}/\fba			       \; & \; 450663(10)      \; & \; 7102.5(6)       \; & \; 752.01(2)       \; & \; 53.290(2)       \; & \; 5.0634(1)       \; & \; 0.240331(7)     \\ 
 \si_{\born}^{\mathrm{var}}/\fba		       \; & \; 446072(10)      \; & \; 6937.6(6)       \; & \; 714.64(2)       \; & \; 48.618(1)       \; & \; 4.4510(1)       \; & \; 0.202315(6)     \\ 
    \hline \hline													  
 \de_{\EW}^{\mu^+\nu_{\mu}\,,\mathrm{var}}/\%	       \; & \; -3.1\phz        \; & \; -5.2\phz        \; & \; -8.2\phz        \; & \; -14.8\phz       \; & \; -22.5\phz       \; & \; -33.0(1)        \\ 
 \de_{\EW}^{\mathrm{rec}\,,\mathrm{var}}/\%	       \; & \; -2.2\phz        \; & \; -3.7\phz        \; & \; -6.8\phz        \; & \; -12.8\phz       \; & \; -19.9\phz       \; & \; -29.4(1)        \\ 
    \hline \hline													  
 \de_{\QCD}^{\mu = \MW}/\%			       \; & \; 47.7(1)         \; & \; 30.5(1)         \; & \; 11.7(1)         \; & \; -15.7(1)        \; & \; -40.6(1)        \; & \; -70.0(1)        \\ 
 \de_{\QCD}^{\mathrm{var}}/\%			       \; & \; 47.6(1)         \; & \; 31.0(1)         \; & \; 14.4(1)         \; & \; -8.8(1)         \; & \; -30.5(1)        \; & \; -56.0(1)        \\ 
    \hline \hline													  
 \de_{\ga,\born}^{\mathrm{var}}/\%		       \; & \; 0.4\phz         \; & \; 0.3\phz         \; & \; 0.4\phz         \; & \; 0.5\phz         \; & \; 0.4\phz         \; & \; 0.3\phz         \\ 
   \hline														  
 \de_{\ga,\NLO}^{\mathrm{var}}/\%		       \; & \; 0.3\phz         \; & \; 0.3\phz         \; & \; 0.4\phz         \; & \; 0.5\phz         \; & \; 0.4\phz         \; & \; 0.3\phz         \\ 
    \hline \hline													  
 \de_{\Int}^{\mathrm{var}}/\%			       \; & \; 0.1\phz        \; & \; 0.01\phantom{()}\; & \; -0.05\phantom{()}\; & \; -0.1\phz       \; & \; -0.1\phz       \; & \; -0.1\phz       \\ 
  \end{array} $$
\mycaption{\label{ta:mtw_LHC} Integrated cross sections for different
  cuts on the transverse mass of the W at the LHC.}
\end{table}

\begin{table}
 																	$$ \begin{array}{c|rrrrrr}
 								    \multicolumn{7}{c}{\Pp\Pp \to \Plp \Pnl\; \mathrm{jet} + X \;\mbox{at} \;\sqrt{s} =14 \TeV} \\
 	       \hline p_{\rT,\mathrm{jet}} / \GeV & 25-\infty \;\;\; & 50-\infty \;\;\; & 100-\infty \;\; & 200-\infty \;\; & 500-\infty \;\; & 1000-\infty \; \\ 
 																		      \hline\hline
 \si_{\born}^{\mu = \MW}/\fba			       \; & \; 508568(11)      \; & \; 182462(4)       \; & \; 49702(1)        \; & \; 8096.3(2)       \; & \; 315.061(5)      \; & \; 11.6750(2)      \\ 
 \si_{\born}^{\mathrm{var}}/\fba		       \; & \; 501826(11)      \; & \; 176106(4)       \; & \; 45313(1)        \; & \; 6488.9(1)       \; & \; 184.742(3)      \; & \; 4.78109(8)      \\ 
    \hline \hline													  
 \de_{\EW}^{\mu^+\nu_{\mu}\,,\mathrm{var}}/\%	       \; & \; -3.0\phz        \; & \; -3.3\phz        \; & \; -4.7\phz        \; & \; -8.6\phz        \; & \; -18.1\phz       \; & \; -28.4(1)        \\ 
 \de_{\EW}^{\mathrm{rec}\,,\mathrm{var}}/\%	       \; & \; -2.2\phz        \; & \; -2.6\phz        \; & \; -4.2\phz        \; & \; -8.3\phz        \; & \; -18.0\phz       \; & \; -28.3(1)        \\ 
    \hline \hline													  
 \de_{\QCD}^{\mu = \MW}/\%			       \; & \; 48.2(1)         \; & \; 64.7(1)         \; & \; 80.6(1)         \; & \; 115.3(1)        \; & \; 188.8(1)        \; & \; 270.2(1)        \\ 
 \de_{\QCD}^{\mathrm{var}}/\%			       \; & \; 47.9(1)         \; & \; 65.5(1)         \; & \; 85.8(1)         \; & \; 135.1(1)        \; & \; 270.3(1)        \; & \; 495.5(1)        \\ 
   \hline														  
 \de_{\QCD,\veto}^{\mu = \MW}/\%		       \; & \; 21.7(1)         \; & \; 18.5(1)         \; & \; 22.5(1)         \; & \; 24.2(1)         \; & \; 5.7(1)	       \; & \; -26.0(1)        \\ 
 \de_{\QCD,\veto}^{\mathrm{var}}/\%		       \; & \; 22.5(1)         \; & \; 21.3(1)         \; & \; 29.9(1)         \; & \; 42.8(1)         \; & \; 52.7(1)         \; & \; 59.5(1)         \\ 
    \hline \hline													  
 \de_{\ga,\born}^{\mathrm{var}}/\%		       \; & \; 0.4\phz         \; & \; 0.7\phz         \; & \; 1.3\phz         \; & \; 2.0\phz         \; & \; 3.4\phz         \; & \; 5.2\phz         \\ 
   \hline														  
 \de_{\ga,\NLO}^{\mathrm{var}}/\%		       \; & \; 0.4\phz         \; & \; 0.7\phz         \; & \; 1.2\phz         \; & \; 1.9\phz         \; & \; 3.3\phz         \; & \; 5.2\phz         \\ 
   \hline														  
 \de_{\ga,\NLO,\veto}^{\mathrm{var}}/\% 	       \; & \; 0.3\phz         \; & \; 0.6\phz         \; & \; 1.1\phz         \; & \; 1.8\phz         \; & \; 3.1\phz         \; & \; 4.7\phz         \\ 
    \hline \hline													  
 \de_{\Int}^{\mathrm{var}}/\%			       \; & \; 0.05\phantom{()}\; & \; 0.1\phz        \; & \; 0.5\phz        \; & \; 1.9\phz        \; & \; 11.5\phz        \; & \; 49.9(1)        \\ 
   \hline														  
 \de_{\Int,\veto}^{\mathrm{var}}/\%		       \; & \; 0.01\phantom{()}\; & \; 0.03\phantom{()}\; & \; 0.1\phz        \; & \; 0.4\phz        \; & \; 1.6\phz        \; & \; 4.7\phz         \\ 
  \end{array} $$
\mycaption{\label{ta:ptj_LHC} Integrated cross sections for different
  cuts on the $p_{\rT}$ of the leading jet at the LHC. Corrections
  with a second jet in real emission events are shown with and without
  a jet veto.}
\end{table}

\reftas{ta:mtw_LHC} and \ref{ta:ptj_LHC} show the analoguous results for 
a variation of cuts on the transverse mass of the final-state leptons, 
defined by
\begin{equation}
\label{eq:MTW}
M_{\rT,\Pl \nu_\Pl} = \sqrt{2 p_{\rT,\Pl} \dsl{p}_{\rT} (1-\cos \phi_{\Pl \nu_\Pl})} 
\, ,
\end{equation}
and the $p_{\rT}$ of the leading jet $p_{\rT,\mathrm{jet}}$,
respectively. The transverse mass and the $p_{\rT,\Pl}$
distributions are particularly relevant for the measurement of the
\PW-boson mass at hadron colliders. For this measurement, \PW-boson
events without or with very little additional jet activity are
selected. Nevertheless, the calculation of the EW corrections in the
presence of an additional jet supplies a handle to quantify how well
the interplay of QCD and EW corrections is understood.

Note that for a given $p_{\rT,\mathrm{jet}}$ both leptons share the
recoil since they stem from a boosted \PW\ boson and are therefore
preferably emitted in the same direction. Accordingly, the LO cross
section for a given cut on $p_{\rT,\Pl}$ is smaller than for the
same cut on $p_{\rT,\mathrm{jet}}$, because on average the required CM
energy is larger. In other words, there is a kinematic and an
additional PDF suppression of events with a cut on the lepton
$p_{\rT}$ compared to events with the same cut on the jet $p_{\rT}$.
For large values of $M_{\rT,\Pl \nu_\Pl}$, the \PW\ boson is
necessarily produced far off shell so that the cross section is
further suppressed. The presented cut values for $M_{\rT,\Pl \nu_\Pl}$
are chosen because one finds $M_{\rT,\Pl \nu_\Pl} = 2 p_{\rT,\Pl}$ for
back-to-back leptons in the rest frame of the decaying \PW\ boson.

For the most inclusive cross section (left columns in \refta{ta:ptl_LHC}
or \refta{ta:ptj_LHC}) the EW corrections are at the
percent level and negative. The difference in scale choice is not
important, and due to the recombination procedure
$\delta^{\mathrm{rec}}_\mathrm{EW}$ is slightly smaller in absolute
size. With increasing $p_{\rT,\Pl}$ cut, the relevant CM energies
rise, and the well-known Sudakov logarithms in the virtual EW
corrections start to dominate the total corrections as expected. For
$p_{\rT,\Pl} > 1000$ GeV, the EW corrections reach the level of
$-30\%$. This behaviour is generic and also holds true for the cross
sections with varying cuts on the transverse mass or
$p_{\rT,\mathrm{jet}}$. 
We compare the EW corrections for $p_{\rT,\mathrm{jet}}$ with 
previous results obtained in an on-shell approximation together 
with the differential distributions in \refse{se:results_mom_distr}.

Turning to the NLO QCD results, the corrections
$\delta_\mathrm{QCD}^\mathrm{var}$ for different cuts on
$p_{\rT,\Pl}$, as shown in \refta{ta:ptl_LHC}, are sizable and reach
the 50\% level for intermediate cut values. For low cut values,
$\delta_\mathrm{QCD}^{\mu=\MW}$ is practically the same. However, for
large cut values, the corrections for a fixed scale differ
significantly. Here, $\delta_\mathrm{QCD}^{\mu=\MW}$ grows large and
negative to compensate for the overestimated LO cross section, which
is larger by more than a factor of two with respect to
$\sigma_0^\mathrm{var}$. This is expected, since the hard jet
recoiling against the high-$p_{\rT}$ lepton should be reflected in the
scale choice. Including the NLO QCD corrections, 
the difference between the results obtained with our two scale choices 
is significantly reduced.

For small cut values on the transverse mass, as
shown in \refta{ta:mtw_LHC}, 
the corrections are quite similar to the ones of the corresponding cuts on
$p_{\rT,\Pl}$.
However, for large $M_{\rT,\Pl \nu_\Pl}$, both scale choices
fail to reflect the kinematical situation, since the production of
a far off-shell \PW\ boson is dominated by the region near the 
threshold set by the cut on $M_{\rT,\Pl \nu_\Pl}$.
In this region the \PW\ boson
decays mainly to back-to-back leptons with relatively soft
jet activity. Hence, $\delta_\mathrm{QCD}^{\mu=\MW}$ as well as
$\de_\QCD^\mathrm{var}$ become large and negative. A scale choice based 
on the CM energy of the event or a scale choice reflecting the invariant 
mass of the lepton pair\footnote{The scale 
$\mu=\sqrt{M_{\Pl\Pnl}^2+(p_{\rT}^\mathrm{W})^2}$, where $M_{\Pl\Pnl}$ is the
invariant mass of the two leptons in the final state and $p_{\rT}^\mathrm{W}$ 
denotes their transverse momentum, would also be an adequate choice.} 
would be more adequate.

As discussed above, the integrated NLO QCD cross sections for large
$p_{\rT,\mathrm{jet}}$ cuts, as shown in \refta{ta:ptj_LHC}, contain
large contributions from a completely different class of events for
which two jets recoil against each other. Hence, the corrections are
huge.  The correction $\delta_\mathrm{QCD}^{\mu=\MW}$ is smaller than
$\delta_\mathrm{QCD}^{\mathrm{var}}$ because it is defined relative to
a larger LO cross section. In absolute size, they are similar. Using the
jet veto proposed at the end of \refse{se:basic_cuts}, the corrections
are reduced and $\delta_\mathrm{QCD}^{\mathrm{var}}$ rises only to the
50\% level for large cut values. The fixed scale choice leads to even
smaller corrections $\delta_\mathrm{QCD}^{\mu=\MW}$ in absolute size.
However, varying the exact definition of the jet veto, the variable
scale turns out to be more robust. We have also verified,
that this simple jet veto indeed removes mainly events with back-to-back jets
from the event selection. If we only veto events with
$\cos\phi_{jj}<-0.99$, where $\phi_{jj}$ is the azimuthal angle in the
transverse plane between the two jets, $\de_{\QCD}^{\mathrm{var}}$ for example is still
reduced from 495\% to 172\% for $p_{\rT,\mathrm{jet}} > 1000$ GeV. Events with 
$\cos\phi_{jj}>0$ do not have any noticeable effect.

The contribution $\delta_\gamma$ from the photon-induced processes are
small and only reach up to 5\% for large cuts on $p_{\rT,\Pl}$ or
$p_{\rT,\mathrm{jet}}$ where the EW and QCD corrections to the
dominating tree processes are by far larger. The NLO corrections to
the photon-induced processes are phenomenologically completely
irrelevant.

The corrections due to the interference between EW and QCD diagrams
also turn out to be unimportant.  They only increase together with the
NLO QCD corrections for large $p_{\rT,\mathrm{jet}}$. Once a sensible
jet veto is applied, they disappear again.

\begin{table}
 																	$$ \begin{array}{c|rrrrrr}
 							      \multicolumn{7}{c}{\Pp\bar\Pp \to \Plp \Pnl\; \mathrm{jet} + X \;\mbox{at} \;\sqrt{s} =1.96 \TeV} \\
 		      \hline p_{\rT,\Pl} / \GeV & 25-\infty \;\;\; & 50-\infty \;\;\; & 75-\infty \;\;\; & 100-\infty \;\; & 200-\infty \;\; & 300-\infty \;\; \\ 
 																		      \hline\hline
 \si_{\born}^{\mu = \MW}/\fba			       \; & \; 37341.5(7)      \; & \; 10560.8(4)      \; & \; 1007.54(4)      \; & \; 263.50(1)       \; & \; 7.2415(4)       \; & \; 0.39000(2)      \\ 
 \si_{\born}^{\mathrm{var}}/\fba		       \; & \; 36056.0(7)      \; & \; 10049.9(4)      \; & \; 863.34(4)       \; & \; 209.484(9)      \; & \; 4.8338(2)       \; & \; 0.23655(1)      \\ 
    \hline \hline													  
 \de_{\EW}^{\mu^+\nu_{\mu}\,,\mathrm{var}}/\%	       \; & \; -2.8\phz        \; & \; -5.4\phz        \; & \; -6.8\phz        \; & \; -8.2\phz        \; & \; -13.2(1)        \; & \; -17.4(1)        \\ 
 \de_{\EW}^{\mathrm{rec}\,,\mathrm{var}}/\%	       \; & \; -1.9\phz        \; & \; -2.9\phz        \; & \; -4.0\phz        \; & \; -5.3(1)         \; & \; -9.1\phz        \; & \; -12.4\phz       \\ 
    \hline \hline													  
 \de_{\QCD}^{\mu = \MW}/\%			       \; & \; 33.5(1)         \; & \; 23.8(1)         \; & \; 27.7(1)         \; & \; 18.3(1)         \; & \; -6.4(1)         \; & \; -22.2(1)        \\ 
 \de_{\QCD}^{\mathrm{var}}/\%			       \; & \; 36.3(1)         \; & \; 27.3(1)         \; & \; 40.0(1)         \; & \; 36.8(1)         \; & \; 28.2(1)         \; & \; 21.5(1)         \\ 
    \hline \hline													  
 \de_{\ga,\born}^{\mathrm{var}}/\%		       \; & \; 0.4\phz         \; & \; 0.5\phz         \; & \; 1.2\phz         \; & \; 1.4\phz         \; & \; 1.5\phz         \; & \; 1.3\phz         \\ 
   \hline														  
 \de_{\ga,\NLO}^{\mathrm{var}}/\%		       \; & \; 0.4\phz         \; & \; 0.5\phz         \; & \; 1.2\phz         \; & \; 1.4\phz         \; & \; 1.5\phz         \; & \; 1.3\phz         \\ 
    \hline \hline													  
 \de_{\Int}^{\mathrm{var}}/\%			       \; & \; -0.1\phz       \; & \; -0.1\phz       \; & \; -0.2\phz       \; & \; -0.2\phz       \; & \; -0.2\phz        \; & \; -0.1\phz        \\ 
  \end{array} $$
\mycaption{\label{ta:ptl_TEV} Integrated cross sections for different
  cuts on the lepton transverse momentum at the Tevatron. }
\end{table}

\begin{table}
 																	$$ \begin{array}{c|rrrrrr}
 							      \multicolumn{7}{c}{\Pp\bar\Pp \to \Plp \Pnl\; \mathrm{jet} + X \;\mbox{at} \;\sqrt{s} =1.96 \TeV} \\
 		    \hline M_{\rT,\Pl\Pnl} / \GeV & 50-\infty \;\;\; & 100-\infty \;\; & 150-\infty \;\; & 200-\infty \;\; & 400-\infty \;\; & 600-\infty \;\; \\ 
 																		      \hline\hline
 \si_{\born}^{\mu = \MW}/\fba			       \; & \; 34421.6(6)      \; & \; 434.45(3)       \; & \; 80.338(2)       \; & \; 27.5868(9)      \; & \; 1.25355(4)      \; & \; 0.088241(3)     \\ 
 \si_{\born}^{\mathrm{var}}/\fba		       \; & \; 33359.8(6)      \; & \; 415.57(3)       \; & \; 75.995(2)       \; & \; 25.9198(8)      \; & \; 1.15703(4)      \; & \; 0.080524(3)     \\ 
    \hline \hline													  
 \de_{\EW}^{\mu^+\nu_{\mu}\,,\mathrm{var}}/\%	       \; & \; -2.9\phz        \; & \; -5.0\phz        \; & \; -6.5\phz        \; & \; -8.0\phz        \; & \; -12.7\phz       \; & \; -16.8\phz       \\ 
 \de_{\EW}^{\mathrm{rec}\,,\mathrm{var}}/\%	       \; & \; -1.9\phz        \; & \; -3.4\phz        \; & \; -4.9\phz        \; & \; -6.2\phz        \; & \; -10.1\phz       \; & \; -13.3\phz       \\ 
    \hline \hline													  
 \de_{\QCD}^{\mu = \MW}/\%			       \; & \; 33.8(1)         \; & \; 20.8(1)         \; & \; 12.5(1)         \; & \; 7.4(1)	       \; & \; -4.7(1)         \; & \; -13.2(1)        \\ 
 \de_{\QCD}^{\mathrm{var}}/\%			       \; & \; 36.2(1)         \; & \; 24.5(1)         \; & \; 16.9(1)         \; & \; 12.4(1)         \; & \; 1.7(1)	       \; & \; -5.9(1)         \\ 
    \hline \hline													  
 \de_{\ga,\born}^{\mathrm{var}}/\%		       \; & \; 0.4\phz         \; & \; 0.2\phz         \; & \; 0.2\phz         \; & \; 0.1\phz         \; & \; 0.1\phz         \; & \; 0.05\phantom{()}\\ 
   \hline														  
 \de_{\ga,\NLO}^{\mathrm{var}}/\%		       \; & \; 0.3\phz         \; & \; 0.2\phz         \; & \; 0.2\phz         \; & \; 0.1\phz         \; & \; 0.1\phz         \; & \; 0.05\phantom{()} \\ 
    \hline \hline													  
 \de_{\Int}^{\mathrm{var}}/\%			       \; & \; -0.1\phz       \; & \; -0.1\phz       \; & \; -0.1\phz       \; & \; -0.1\phz       \; & \; -0.1\phz       \; & \; -0.04\phantom{()}\\ 
  \end{array} $$
\vspace{-0.7cm}
\mycaption{\label{ta:mtw_TEV} Integrated cross sections for different
  cuts on the transverse mass of the W at the Tevatron.}
\end{table}

\begin{table}
 																	$$ \begin{array}{c|rrrrrr}
 							      \multicolumn{7}{c}{\Pp\bar\Pp \to \Plp \Pnl\; \mathrm{jet} + X \;\mbox{at} \;\sqrt{s} =1.96 \TeV} \\
 	     \hline p_{\rT,\mathrm{jet}} / \GeV & 25-\infty \;\;\; & 50-\infty \;\;\; & 75-\infty \;\;\; & 100-\infty \;\; & 200-\infty \;\; & 300-\infty \;\; \\ 
 																		      \hline\hline
 \si_{\born}^{\mu = \MW}/\fba			       \; & \; 37341.5(7)      \; & \; 8848.7(2)       \; & \; 3115.31(6)      \; & \; 1231.80(2)      \; & \; 54.5590(8)      \; & \; 3.62805(6)      \\ 
 \si_{\born}^{\mathrm{var}}/\fba		       \; & \; 36056.0(7)      \; & \; 8094.1(2)       \; & \; 2686.10(5)      \; & \; 998.31(2)       \; & \; 34.9921(6)      \; & \; 1.89648(3)      \\ 
    \hline \hline													  
 \de_{\EW}^{\mu^+\nu_{\mu}\,,\mathrm{var}}/\%	       \; & \; -2.8\phz        \; & \; -2.9\phz        \; & \; -3.2\phz        \; & \; -3.7\phz        \; & \; -6.5\phz        \; & \; -9.2\phz        \\ 
 \de_{\EW}^{\mathrm{rec}\,,\mathrm{var}}/\%	       \; & \; -1.9\phz        \; & \; -2.2\phz        \; & \; -2.6\phz        \; & \; -3.2\phz        \; & \; -6.2\phz        \; & \; -9.0\phz        \\ 
    \hline \hline													  
 \de_{\QCD}^{\mu = \MW}/\%			       \; & \; 33.5(1)         \; & \; 31.7(1)         \; & \; 25.1(1)         \; & \; 20.7(1)         \; & \; 6.2(1)	       \; & \; -8.7(1)         \\ 
 \de_{\QCD}^{\mathrm{var}}/\%			       \; & \; 36.3(1)         \; & \; 39.5(1)         \; & \; 39.5(1)         \; & \; 41.9(1)         \; & \; 56.6(1)         \; & \; 70.5(1)         \\ 
   \hline														  
 \de_{\QCD,\veto}^{\mu = \MW}/\%		       \; & \; 20.9(1)         \; & \; 7.4(1)	       \; & \; 1.4(1)	       \; & \; -3.7(1)         \; & \; -24.0(1)        \; & \; -43.8(1)        \\ 
 \de_{\QCD,\veto}^{\mathrm{var}}/\%		       \; & \; 24.0(1)         \; & \; 15.4(1)         \; & \; 15.2(1)         \; & \; 16.4(1)         \; & \; 19.7(1)         \; & \; 21.1(1)         \\ 
    \hline \hline													  
 \de_{\ga,\born}^{\mathrm{var}}/\%		       \; & \; 0.4\phz         \; & \; 0.8\phz         \; & \; 1.2\phz         \; & \; 1.4\phz         \; & \; 2.0\phz         \; & \; 2.3\phz         \\ 
   \hline														  
 \de_{\ga,\NLO}^{\mathrm{var}}/\%		       \; & \; 0.4\phz         \; & \; 0.8\phz         \; & \; 1.1\phz         \; & \; 1.3\phz         \; & \; 1.8\phz         \; & \; 2.2\phz         \\ 
   \hline														  
 \de_{\ga,\NLO,\veto}^{\mathrm{var}}/\% 	       \; & \; 0.4\phz         \; & \; 0.7\phz         \; & \; 1.0\phz         \; & \; 1.3\phz         \; & \; 1.8\phz         \; & \; 2.1\phz         \\ 
    \hline \hline													  
 \de_{\Int}^{\mathrm{var}}/\%			       \; & \; -0.1\phz       \; & \; -0.4\phz       \; & \; -0.6\phz       \; & \; -0.8\phz       \; & \; -2.0\phz        \; & \; -3.5\phz        \\ 
   \hline														  
 \de_{\Int,\veto}^{\mathrm{var}}/\%		       \; & \; -0.04\phantom{()}\; & \; -0.1\phz       \; & \; -0.3\phz       \; & \; -0.4\phz       \; & \; -0.7\phz       \; & \; -1.1\phz       \\ 
  \end{array} $$
\vspace{-0.7cm}
\mycaption{\label{ta:ptj_TEV} Integrated cross sections for different
  cuts on the $p_{\rT}$ of the leading jet at the Tevatron. Corrections
  with a second jet in real emission events are shown with and without
  a jet veto.}
\end{table}

The qualitative features of the corrections at the Tevatron, where
protons and antiprotons collide at 
$\sqrt{s}=1.96\TeV$, are very similar to those at the LHC. At the
Tevatron the high-energy, Sudakov regime is not as accessible as at
the LHC but the onset of the Sudakov dominance is nevertheless visible
as can be seen for the different observables in 
\reftas{ta:ptl_TEV}--\ref{ta:ptj_TEV}. We have adapted the range
for the different integrated cross sections to the kinematic reach of
the Tevatron.

\subsection{Results on momentum and transverse-mass distributions}
\label{se:results_mom_distr}

In \reffis{fi:pt_all}--\ref{fi:philnu_all} we show for various
observables the LO distribution and the distribution including the
full set of corrections, \ie EW corrections $\delta_{\EW}$, the
contribution of the photon-induced processes $\de_{\ga,\mathrm{NLO}}$, 
interference 
contribution $\de_{\Int}$, and the QCD corrections. The various
contributions to the corrections are also shown separately relative to
the LO.

While the corrections to the integrated cross sections are quite
similar for a given $p_{\rT,\Pl}$ and an $M_{\rT,\Pl \nu_\Pl}$ cut of
similar size, the differential distributions in \reffi{fi:pt_all} and
\reffi{fi:mt_all} are significantly different. The EW corrections for the
$M_{\rT,\Pl \nu_\Pl}$ distributions resemble the corrections for the
inclusive \PW-boson sample for which no additional jet is required
(see, e.g., Figure~2 in \citere{Brensing:2007qm}). This result is
expected since the definition \refeq{eq:MTW} of the transverse mass is
boost invariant to first order in the boost velocity
and therefore insensitive to a boost of
the intermediate \PW\ boson. The $p_{\rT,\Pl}$ distribution, in
contrast, is sensitive to these boosts, and neither the LO prediction
nor the NLO EW corrections resemble the inclusive result (see, e.g.,
Figure~1 in \citere{Brensing:2007qm}).

\bfi     
\bce
\includegraphics[width=15.7cm]{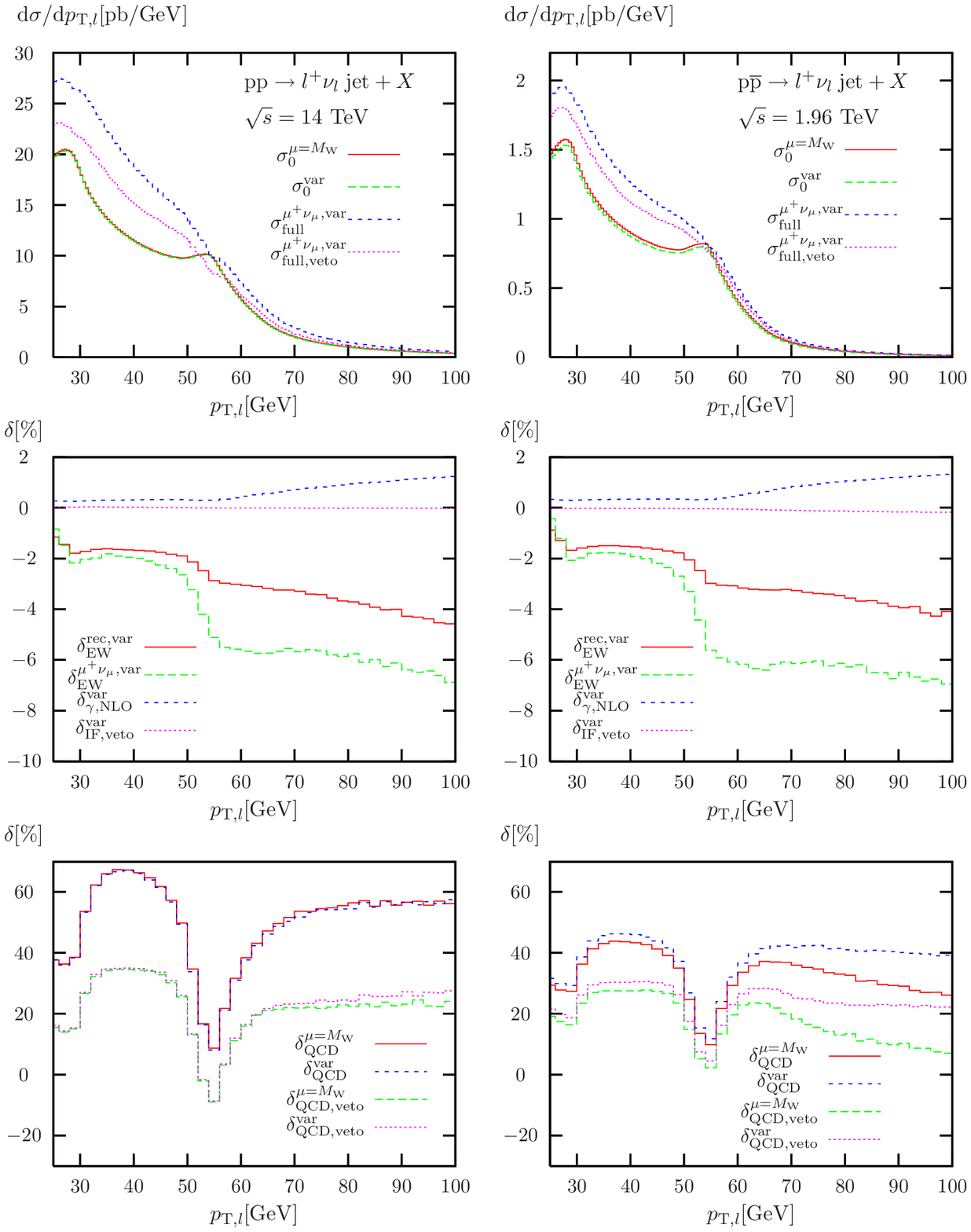} 
\ece
\mycaption{\label{fi:pt_all} LO and fully corrected distribution (top),
  corresponding relative EW, photon-induced, and interference corrections (middle), and 
  relative QCD corrections (bottom) for the transverse momentum
  of the charged lepton at the LHC (left) and the Tevatron (right).
  }
\efi 

\bfi     
\bce
\includegraphics[width=15.7cm]{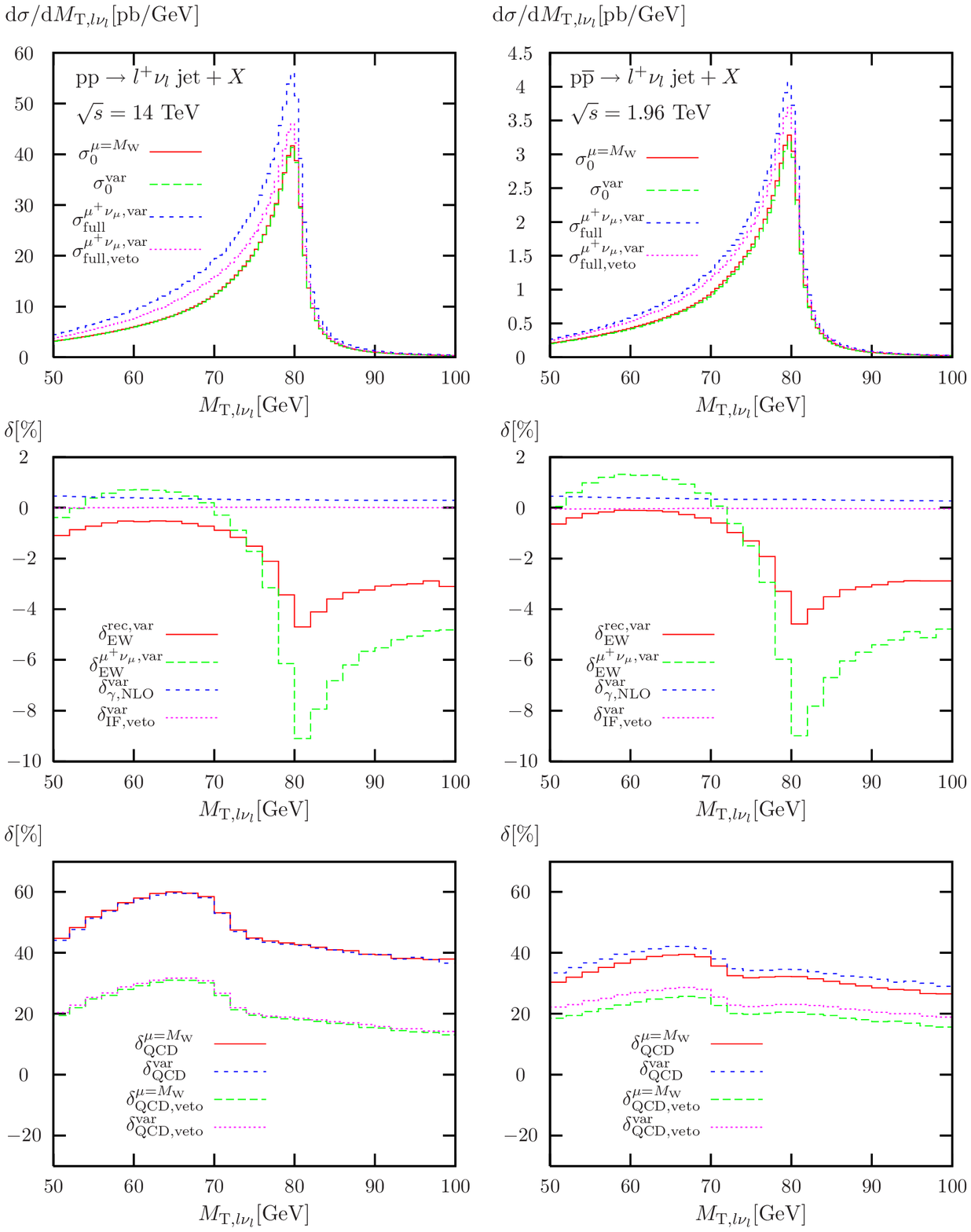} 
\ece
\mycaption{\label{fi:mt_all} LO and fully corrected distribution (top),
  corresponding relative EW, photon-induced, and interference corrections (middle), and 
  relative QCD corrections (bottom) for the 
  \PW\ transverse mass at the LHC (left) and the Tevatron (right).
  }
\efi 

As expected, the corrections for bare muons are larger since photons,
being radiated collinear\-ly to the charged lepton, carry away
transverse momentum. Hence, events that are enhanced by muon-mass
logarithms are shifted to lower bins in the distributions and
to some extent do not survive the basic cuts. As a result, the corrections
are dominated by negative virtual corrections that are not
compensated by positive bremsstrahlung contributions. This is particularly 
evident around the peak of the differential cross section with respect
to the \PW-boson transverse mass in \reffi{fi:mt_all} and also for 
the peaks 
in the transverse-momentum distributions of the charged lepton near
$(\MW\pm p^{\mathrm{cut}}_{\rT,\mathrm{jet}})/2$ in \reffi{fi:pt_all}.

\bfi     
\bce
\includegraphics[width=15.7cm]{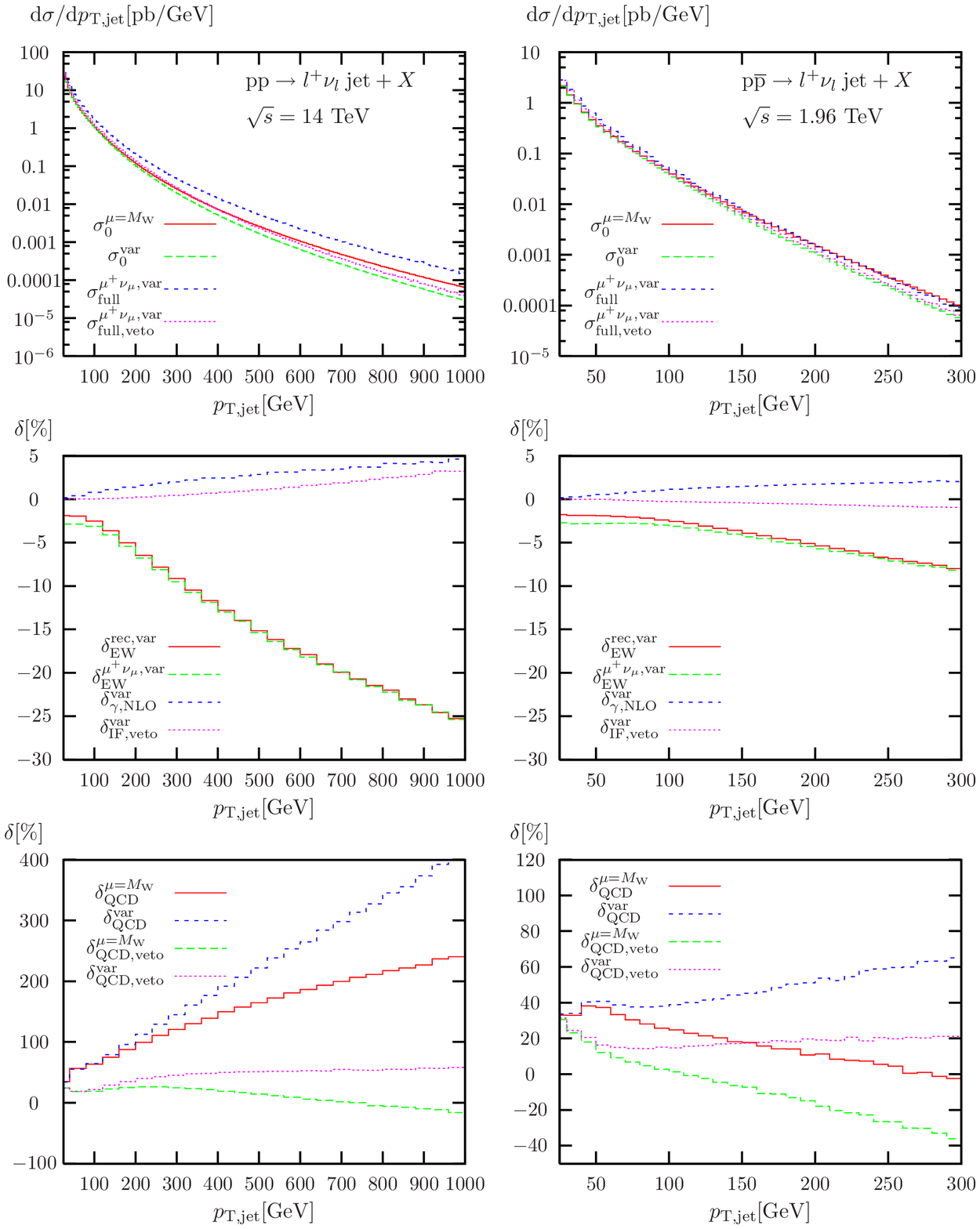} 
\ece
\mycaption{\label{fi:ptj_all} LO and fully corrected distribution (top),
  corresponding relative EW, photon-induced, and interference corrections (middle), and 
  relative QCD corrections (bottom) for the the transverse momentum
  of the leading jet at the LHC (left) and the Tevatron (right).
  }
\efi 

In \reffi{fi:ptj_all} we show the differential cross sections with
respect to $p_{\rT,\mathrm{jet}}$ and the corresponding corrections.
As expected, the increasing size of the EW corrections with
$p_{\rT,\mathrm{jet}}$ due to the EW Sudakov logarithms can be
observed. This observable has also been accessible in
calculations using the approximations of a stable, on-shell \PW\
boson. A comparison of our numerical results to former results for
on-shell $\PW+\mathrm{jet}$ production \cite{Kuhn:2007qc,Kuhn:2007cv,Hollik:2007sq}
has to face the problem that we apply various event-selection cuts to
the leptonic final state, while in the previous papers the degrees of
freedom related to the decaying W are implicitly integrated
out. Nevertheless, the relative EW corrections at high momentum
transfer are dominated by Sudakov logarithms of the form
$\ln^2(\hat{s}/M_{\PW}^2)$ that, at least at the one-loop level, give
rise to large process-independent contributions and therefore are
expected to show a similar behaviour for both the on- and off-shell
case. Comparing our results for the leading-jet $p_{\rT,\mathrm{jet}}$
(\reffi{fi:ptj_all}) with Fig.\ 5 in \citere{Kuhn:2007cv}, we in fact
find agreement within 2\%.
The results for the EW corrections to the integrated cross sections with
different cuts on $p_{\mathrm{T,jet}}$ given in \refta{ta:ptj_LHC} 
also agree within 1\% with the results presented in Fig.\ 9(b) of 
\citere{Hollik:2007sq} for cut values larger than 200 GeV. This figure shows the
relative corrections for the average of $\PWp$- and $\PWm$ production
at the LHC, but the relative EW
corrections to the on-shell $\PWp$ and $\PWm$ production rates turn
out to be very similar, as can e.g.\ be seen in Fig.\ 10 of \citere{Kuhn:2007cv}. 
Comparing the EW corrections at the Tevatron given in Table 6 to
the on-shell results of Fig.\ 9(a) in \citere{Hollik:2007sq}, we observe slightly
larger deviations, because the universal Sudakov-like contributions
are not dominant at typical Tevatron energy scales.

Turning again to the NLO QCD results, the corrections to
$p_{\rT,\Pl}$ and $M_{\rT,\Pl \nu_\Pl}$
distributions, also displayed in \reffis{fi:pt_all} and \ref{fi:mt_all}, show
quite different features. The corrections to the $M_{\rT,\Pl \nu_\Pl}$
distribution are flatter, 
reflecting the well-known fact that the transverse mass is less sensitive to
additional QCD radiation. In contrast, the corrections
$\delta_\mathrm{QCD}$ for $p_{\rT,\Pl}$ show pronounced dips where
the LO cross section has peaks.  The real corrections do not
particularly populate the regions of the distributions that are
enhanced due to the particular LO kinematics.
The QCD corrections to the 
differential distribution for $p_{\rT,\mathrm{jet}}$ show
exactly the same features which have already been discussed for the
integrated cross sections (see \refta{ta:ptj_LHC}), as can be seen 
in \reffi{fi:ptj_all}. 

At the Tevatron, 
the shapes of the EW and QCD corrections to distributions (see
\reffis{fi:pt_all}--\ref{fi:ptj_all}) are very similar
to the respective results for the 
LHC. For the $p_\rT$ distribution of the leading jet
(see \reffi{fi:ptj_all}), the jet veto again stabilizes the
perturbative result. However, using the variable scale choice, the
increase in cross section without jet veto is not as pronounced as at
the LHC.  On the other hand, as expected, the fixed scale choice
together with a jet veto leads to large negative corrections. A fixed
scale choice without a jet veto accidentally leads to small
corrections at the Tevatron.

\subsection{Results on rapidity and angular distributions}

In \reffi{fi:yl_all}, we analyse the rapidity distribution for the
charged lepton. While the EW corrections are flat, the NLO QCD
corrections are larger at large rapidities and, hence, tend to
populate the forward and backward regions with more events. Concerning
the rapidity of the leading jet at the LHC, both EW and NLO QCD corrections do
not disturb the LO shapes of the distribution, as can be seen in
\reffi{fi:yj_all}.

At the Tevatron, the rapidity distributions show the expected
asymmetry between the forward and backward direction due to the
antiproton in the initial state. This asymmetry is also reflected by
asymmetric NLO QCD corrections for the rapidity of the leading jet.

\bfi     
\bce
\includegraphics[width=15.7cm]{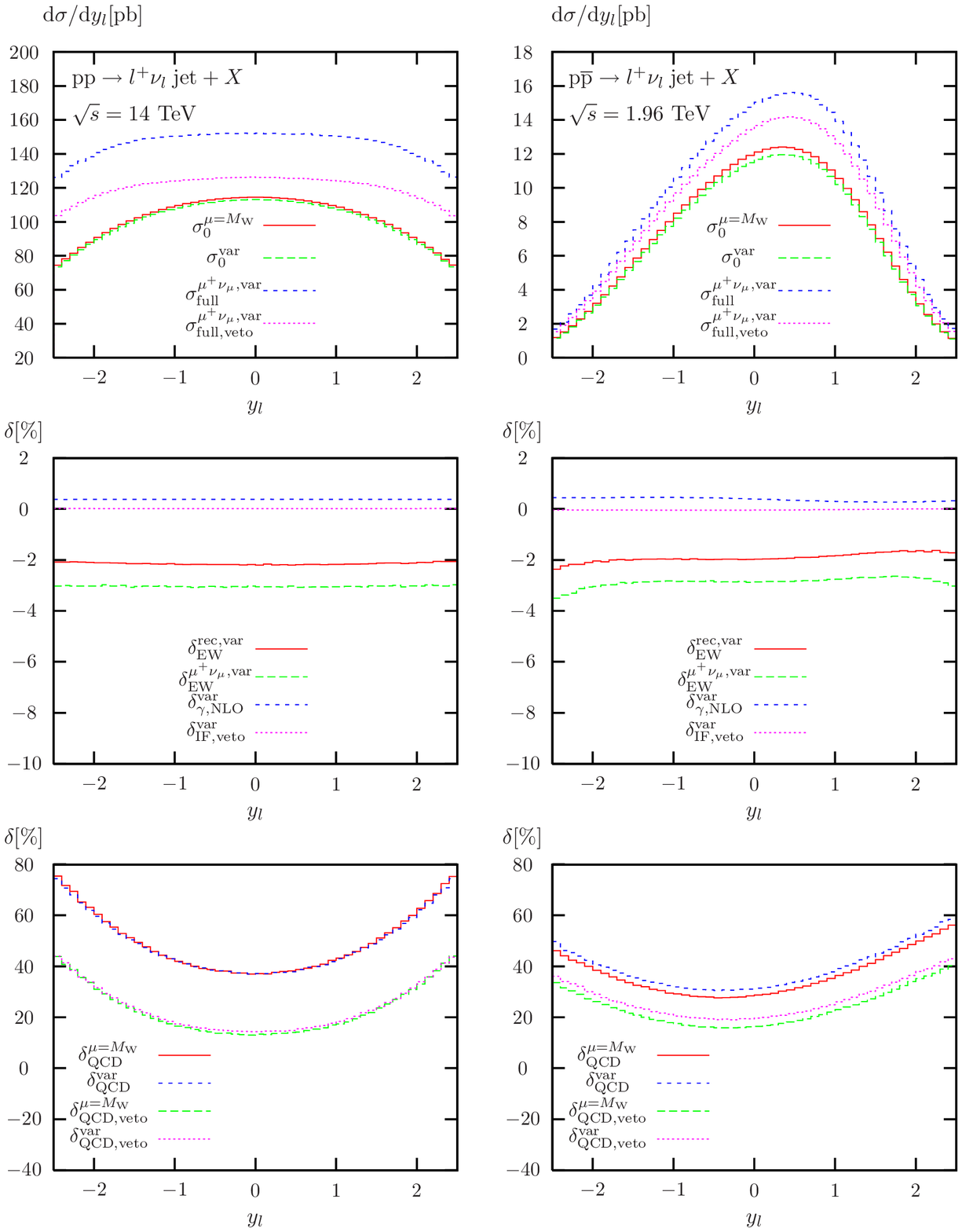} 
\ece
\mycaption{\label{fi:yl_all} LO and fully corrected distribution (top),
  corresponding relative EW, photon-induced, and interference corrections (middle), and 
  relative QCD corrections (bottom) for the rapidity
  of the charged lepton at the LHC (left) and the Tevatron (right).
  }
\efi 

\bfi     
\bce
\includegraphics[width=15.7cm]{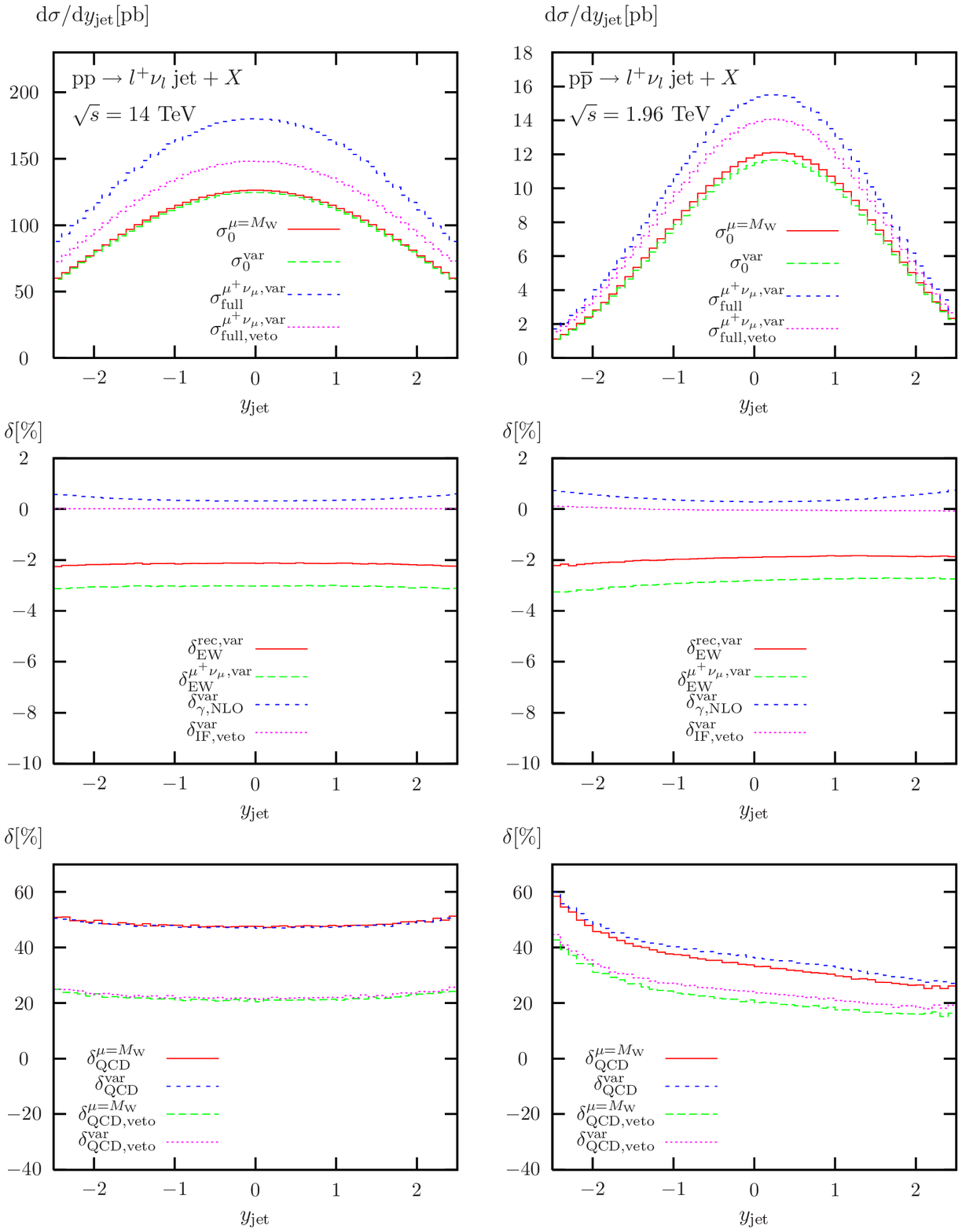} 
\ece
\mycaption{\label{fi:yj_all} LO and fully corrected distribution (top),
  corresponding relative EW, photon-induced, and interference corrections (middle), and 
  relative QCD corrections (bottom) for the rapidity
  of the leading jet at the LHC (left) and the Tevatron (right).
  }
\efi 

\bfi     
\bce
\includegraphics[width=15.7cm]{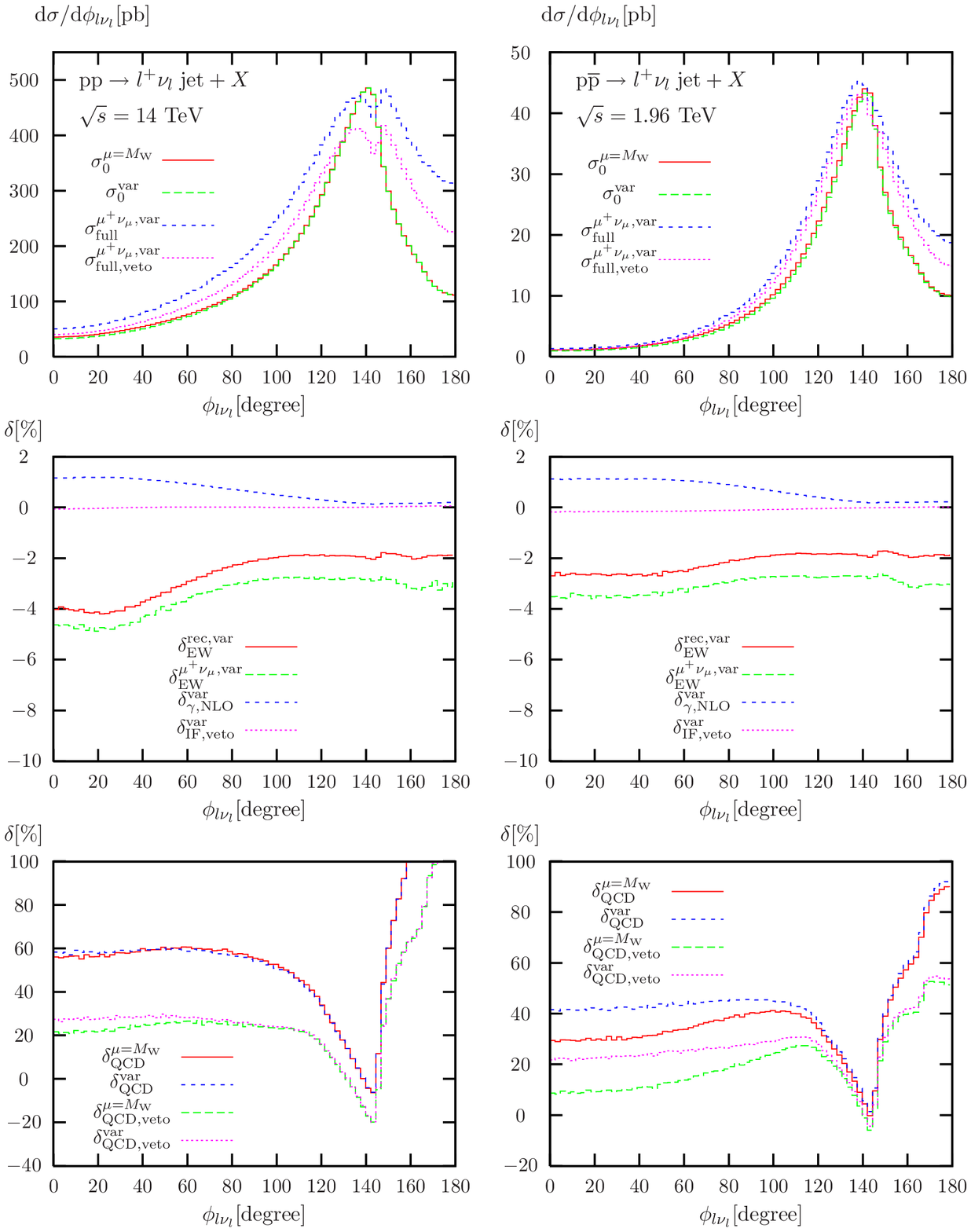} 
\ece
\mycaption{\label{fi:philnu_all} LO and fully corrected distribution (top),
  corresponding relative EW, photon-induced, and interference corrections (middle), and 
  relative QCD corrections (bottom) for the azimuthal angle 
  in the transverse plane between the charged lepton and the 
  neutrino (missing $p_{\rT}$) at the LHC (left) and the Tevatron (right).
  }
\efi 

Another interesting observable is the angle between the charged lepton
and missing $p_{\rT}$ in the transverse plane (\reffi{fi:philnu_all}).
For \PW\ production without jet activity the two leptons are always
back-to-back in the transverse plane. Here, with one jet at LO, the
distribution is still peaked at large angles. However, back-to-back
events are suppressed as shown in \reffi{fi:philnu_all}. While the EW
corrections only slightly disturb the shape of the distribution, the
NLO QCD corrections tend to distribute events more equally with
respect to the investigated angle.  However, the dip in the NLO
distributions at the LO peak might indicate that higher orders are
necessary for an accurate prediction of this observable.
The shapes of the relative QCD corrections reflect the
large impact of real corrections induced by $\PW+2\,\mathrm{jets}$ configurations
where two hard jets are nearly back-to-back while the W~boson 
receives only a small transverse momentum. Such events cause the
large positive corrections for $\phi_{l\nu_l}\to180^\circ$, which
are sensitive to the application of the jet veto.

\section{Conclusions}
\label{se:concl}

We have presented the first calculation of the full electroweak (EW) NLO
corrections for \PW-boson hadroproduction in association with a hard
jet where all off-shell effects are taken into account in the leptonic
\PW-boson decay, i.e.\ we have studied final states with a jet, a
charged lepton, and missing transverse momentum at NLO in the EW coupling
constant within the SM.

We have implemented our results in a flexible Monte Carlo code which
can model the experimental event definition at the NLO
parton level. The distinction of $\PW+\mathrm{jet}$ and
$\PW+\mathrm{photon}$ production is consistently implemented by making
use of the measured quark-to-photon fragmentation function.
We have also recalculated the NLO QCD corrections
supporting a phase-space dependent scale choice.
Interference contributions by EW and QCD diagrams
as well as photon-induced processes,
contributing at the same order, are included but phenomenologically
unimportant. 

The presented integrated cross sections and differential distributions
demonstrate the applicability of our calculation.
The EW corrections to the transverse mass of the \PW\ boson exhibit the
same enhancement as for \PW\ bosons without jet activity, reaching 
$-10\%$ at the peak of the LO distribution which is dominated by
resonant \PW\ bosons. For large transverse mass, i.e.\ in the off-shell
tail of the distribution, we find large negative corrections, dominated
by the well-known EW Sudakov logarithms.  The EW corrections to the
$p_\rT$ distributions of the final-state particles are rather flat and
at the percent level for small values  of $p_\rT$ and also become more
and more negative owing to contributions from Sudakov logarithms in
accordance with previous on-shell approximations.  The QCD corrections
have a typical size of 50\%. However, they can  become extremely large
(hundreds of percent) at large jet $p_\rT$  if one does not apply a
sensible jet veto.

The precise prediction for \PW-boson production at the Tevatron and the
LHC is an important task. Our results extend the theoretical effort  to
associated production with a hard jet. As part of a full NNLO
prediction of the mixed EW and QCD corrections for inclusive  \PW\
production our results can provide a handle for a better 
understanding of the interplay between EW and QCD corrections in the 
charged-current Drell--Yan process. Moreover, they establish a flexible
precision calculation for one of the most important background
processes for new-physics searches.
In the range of intermediate and large transverse momenta of the
additional hard jet our calculation delivers state-of-the-art
predictions, for small transverse momenta the pure NLO calculation
should of course be improved by dedicated QCD resummations, a task that 
goes beyond the scope of this paper.

\subsection*{Acknowledgements}

TK wants to thank Max Huber, Thomas Hahn, and Stefan Kallweit for many
interesting and helpful discussions. This work
is supported in part by the European Community's Marie-Curie Research
Training Network under contract MRTN-CT-2006-035505 ``Tools and
Precision Calculations for Physics Discoveries at Colliders''.

\end{document}